\RequirePackage[2020-02-02]{latexrelease}
\documentclass[%
 reprint,
superscriptaddress,
 amsmath,amssymb,
 aps,
]{revtex4-2}

\usepackage{graphicx}
\usepackage{dcolumn}
\usepackage{bm}
\usepackage{hyperref}

\usepackage[printwatermark]{xwatermark}
\newwatermark[page=1,color=black!100,angle=0,scale=0.32,xpos=70,ypos=132]{NORDITA 2023-006}

\usepackage{xcolor}

\begin{document}

\preprint{APS/123-QED}

\title{Cosmological forecast of the 21-cm power spectrum using the halo model of reionization}

\author{Aurel Schneider}
\email{aurel.schneider@uzh.ch}
\author{Timoth\'ee Schaeffer}%
\affiliation{Institute for Computational Science, University of Zurich, Winterthurerstrasse 190, 8057 Zurich, Switzerland.}

\author{Sambit K. Giri}
\affiliation{Nordita, KTH Royal Institute of Technology and University of Stockholm, Hannes Alvens v\"ag 12 SE-106 91 Stockholm, Sweden.}%

\date{\today 
}

\begin{abstract}
The 21-cm power spectrum of reionization is a promising probe for cosmology and fundamental physics. Exploiting this new observable, however, requires fast predictors capable of efficiently scanning the very large parameter space of cosmological and astrophysical uncertainties. 
In this paper, we introduce the halo model of reionization ({\tt HMreio}), a new analytical tool that combines the halo model of the cosmic dawn with the excursion-set bubble model for reionization, assuming an empirical correction factor to deal with overlapping ionization bubbles. First, {\tt HMreio} is validated against results from the well-known semi-numerical code {\tt 21cmFAST}, showing a good overall agreement for wave-modes of $k\lesssim 1$ h/Mpc. Based on this result, we perform a Monte-Carlo Markov-Chain (MCMC) forecast analysis assuming mock data from 1000-hour observations with the low-frequency part of the Square Kilometre Array (SKA) observatory. We simultaneously vary the six standard cosmological parameters together with seven astrophysical nuisance parameters quantifying the abundance and spectral properties of sources. Depending on the assumed theory error, we find very competitive constraints on cosmological parameters. In particular, it will be possible to conclusively test current cosmological tensions related to the Hubble parameter ($H_0$-tension) and the matter clustering amplitude ($S_8$-tension). Furthermore, the sum of the neutrino masses can be strongly constrained, making it possible to determine the neutrino mass hierarchy at the $\sim 90$ percent confidence level. However, these goals can only be achieved if the current modelling uncertainties are substantially reduced to below $\sim 3$ percent.
\end{abstract}

\maketitle


\section{Introduction}
The first detection of the 21-cm signal from the epoch of reionization and the cosmic dawn will open up a new window to our universe \cite{Furlanetto:2006jb,Pritchard:2011xb}. The signal will contain information about the distribution of matter as well as the thermal and ionization history of the intergalactic medium (IGM) at redshifts between about 6 and 25, an epoch that has so far been largely unobserved. As a consequence, we are expected to learn more about the properties of the very first stars and galaxies indirectly probing the presence of population III stars \cite{Mirocha:2017xxz,Mebane:2020jwl,Magg:2021jyc}, intermediate-mass black holes \cite{Kulkarni:2017qwu,Ventura:2022rwn}, or radio-emitting sources \cite{Fialkov:2019vnb,Reis:2020arr}. Furthermore, we may find evidence for exotic interactions between the standard model and the dark sector \cite{Lopez-Honorez:2018ipk,Driskell:2022pax,Barkana:2022hko} or constrain different dark matter scenarios \cite[e.g.][]{Lopez-Honorez:2017csg,Schneider:2018xba,Munoz:2020mue,Giri:2022nxq}. 

Next to high-redshift star formation, black holes, or dark matter physics, the 21-cm signal is also sensitive to cosmological parameters of the $\Lambda$CDM model. It has been pointed out nearly two decades ago that the 21-cm power spectrum can be used to constrain the standard cosmological parameters at a level competitive to other cosmological probes \cite{McQuinn:2005hk,Mao:2008ug}. More recent forecast studies have confirmed these conclusions \cite{Liu:2015gaa,Kern:2017ccn}, showing that future 21-cm observations from the Hydrogen Epoch of Reionization Array \citep[HERA,][]{DeBoer:2016tnn} could yield constraints on e.g. the clustering amplitude ($\sigma_8$) or the matter abundance ($\Omega_m$) comparable to the cosmic microwave background (CMB) limits from the {\tt Planck} satellite \cite{Planck:2018vyg}. 

In terms of observations, we are still awaiting the first confirmed detection of the 21-cm signal from the epoch of reionization. So far, ongoing surveys from the Low-Frequency Array \citep[LOFAR,][]{vanHaarlem:2013dsa}, the Murchison Widefield Array \citep[MWA,][]{Tingay:2012ps}, or the Hydrogen Epoch of Reionization Array \cite[HERA,][]{DeBoer:2016tnn} have found upper limits that have been used to exclude some extreme models of reionization \citep{Ghara:2020syx,ghara2021constraining,greig2021interpreting}. However, upcoming observations from HERA and from the low-frequency component of the Square Kilometre Array \cite[SKA-Low,][]{Koopmans:2015sua}, currently under construction in western Australia, are expected to provide the first detection of the 21-cm power spectrum at $z\gtrsim 6$. Such a detection will mark the beginning of a new era in radio astronomy, providing a different and novel way to observe the Universe.

In this paper, we present a new method to predict the 21-cm signal at the epoch of reionization and the cosmic dawn. The method is based on a combination of the bubble size function from \citet{Furlanetto:2004nh} with the halo model approach for the cosmic dawn \citep{Schneider:2020xmf} to obtain very fast predictions for the global signal and the power spectrum at $z\sim6-25$. The halo model approach was originally introduced to calculate the nonlinear clustering signal using information about halo numbers, bias, and profiles \citep{Seljak:2000gq,Peacock:2000qk,Cooray:2002dia}. However, it turns out that the same formalism can be used to describe the distribution of radiation from sources \citep{Holzbauer:2011mv, Schneider:2020xmf} which is required to obtain the 21-cm signal.

As a first step, we describe the halo model of reionization ({\tt HMreio}) and compare it to predictions from the semi-numerical code {\tt 21cmFAST} \cite{Mesinger:2011aaa,Murray:2020dcd}. We then perform a forecast analysis based on mock data from 1000 observing hours with the low-frequency part of the SKA telescope (SKA-Low). We provide estimates for the errors on the standard five cosmological parameters of the $\Lambda$CDM model plus on the sum of the neutrino masses. Along with these parameters, we also investigate the potential of SKA-Low to constrain the standard astrophysical parameters such as the photon escape fraction, the stellar-to-halo mass relation, and the amount of X-ray emission.

The paper is structured as follows: In Sec.~\ref{sec:model} we present the formalism of the {\tt HMreio} model. Sec.~\ref{sec:comparison} is dedicated to the validation of {\tt HMreio} including a detailed comparison with {\tt 21cmFAST}. In Sec.~\ref{sec:forecast} we present a forecast study investigating the posterior contours on both cosmological and astrophysical parameters. A summary of the model and the forecast analysis is provided in Sec.~\ref{sec:conclusions}.

\section{Modelling the 21cm signal}\label{sec:model}
In this section, we show how to predict the 21cm global signal and power spectrum using the halo model of reionization ({\tt HMreio}). The model consists of an extension of \citet{Schneider:2020xmf} and we refer to this work for all components that are not defined here.

The observable 21-cm signal is given by the differential brightness temperature 
\begin{multline}\label{dTb}
dT_{b}(\mathbf{x},z)\simeq T_0(z) x_{\rm HI}(\mathbf{x},z)\left[1+\delta_b(\mathbf{x},z)\right]\\ 
\times\frac{x_{\alpha}(\mathbf{x},z)}{1+x_{\alpha}(\mathbf{x},z)}\left[1-\frac{T_{\rm cmb}(z)}{T_{\rm gas}(\mathbf{x},z)}\right],
\end{multline}
which depends on position ($\mathbf{x}$) and redshift ($z$) \cite[see e.g. Ref.][]{Furlanetto:2006tf}. The position-dependent quantities are the neutral fraction ($x_{\rm HI}$), the baryon overdensity ($\delta_b$), the total UV coupling coefficient ($x_{\alpha}$), and the gas temperature ($T_{\rm gas}$). The background (CMB) temperature ($T_{\rm cmb}$) is assumed to only evolve with redshift. The amplitude of the differential brightness temperature is given by
\begin{eqnarray}
T_0(z)=27 \left(\frac{\Omega_bh^2}{0.023}\right)\left(\frac{0.15}{\Omega_mh^2}\frac{1+z}{10}\right)^{\frac{1}{2}}\,\,\, {\rm mK},
\end{eqnarray}
where $\Omega_m$ and $\Omega_b$ are the cosmic matter and baryon abundances and $h=H_0/100$ (km/s)/Mpc is the dimensionless Hubble parameter.

In the following subsections, we will show how Eq.~(\ref{dTb}) can be separated into individual components in order to obtain the power spectrum and global signal. Each component is computed using the halo-model approach. Specific care needs to be taken when modelling the overlap of reionization bubbles which cannot be naturally achieved within the halo model. As explained in the following, we will apply an empirical correction factor to deal with this issue of overlap.

\subsection{Power spectrum from individual components}\label{sec:PS}
Without loss of generality, the differential brightness temperature (described in Eq.~\ref{dTb}) can be written as $dT_b = {\bar T_{21}}(1+\delta_{21})$, where ${\bar T}_{21}$ is the average background brightness temperature and  $\delta_{21}$ is the 21-cm perturbation field. The latter can be written in terms of its individual components, i.e.,
\begin{multline}\label{delta21}
\delta_{21} = \beta_r\delta_{r} + \beta_b\delta_{b} + \beta_T\delta_{T} + \beta_{\alpha}\delta_{\alpha}\\
+ \beta_r\beta_b\delta_r\delta_b + \beta_r\beta_T\delta_r\delta_T + \beta_r\beta_{\alpha}\delta_r\delta_{\alpha}- \delta_{dv},
\end{multline}
where the $\beta$-factors only depend on redshift. They are given by $\beta_r=-(1-{\bar x}_{\rm HI})/{\bar x}_{\rm HI}$, $\beta_b\simeq 1$, $\beta_T \simeq T_{\rm cmb}/({\bar T}_{\rm gas}-T_{\rm cmb})$, and $\beta_{\alpha}=1/(1+{\bar x}_{\alpha})$, variables with a bar referring to average background quantities. Note that Eq.~(\ref{delta21}) has been linearised with respect to $\delta_b$ (baryon field), $\delta_T$ (temperature field), and $\delta_{\alpha}$ (Lyman-$\alpha$ coupling field), while $\delta_r$ (reionization bubble field) remains nonlinear. Indeed, the individual perturbations are expected to be small except for the case of the ionization field that has fluctuations of order unity \citep{Pritchard:2011xb}. The last term of Eq.~(\ref{delta21}) is caused by the effect of redshift-space distortions. At the linear level, it can be written as $\delta_{dv}=-\mu^2\delta_m$ \cite{Kaiser:1987qv,Bharadwaj:2004nr}, where $\delta_m$ is the matter perturbation field and $\mu$ is the cosine of the angle with respect to the line-of-sight.

From Eq.~(\ref{delta21}), it is straightforward to calculate the decomposed 21-cm power spectrum. At the linear level in all perturbations (including $\delta_r$) the angle-averaged power spectrum can be written as \cite{Barkana:2004zy}
\begin{multline}\label{Plin}
P^{\rm (lin)}_{21} = P_{r,r} + P_{b,b} + P_{T,T} + P_{\alpha,\alpha}\\
 + 2\left(P_{r,b} + P_{r,T} + P_{r,\alpha} + P_{b,T} + P_{b,\alpha} + P_{T,\alpha}\right)\\
+\frac{2}{3}\left(P_{r,m} + P_{b,m} + P_{T,m} + P_{\alpha,m} \right) + \frac{1}{5}P_{m,m}\,,
\end{multline}
where the individual terms are written as $P_{X,Y}=\beta_X\beta_Y\langle\delta_X,\delta_Y\rangle$ with $X,Y=\lbrace r,b,T,\alpha\rbrace$. It has been pointed out early on \citep[e.g. Refs.][]{Lidz:2006vj,Georgiev:2021yvq} that the fluctuations in the ionization field are of order unity and should therefore not be linearised. We therefore also keep terms that are nonlinear in $\delta_r$, i.e., 
\begin{multline}\label{Pnl}
P^{\rm (nl)}_{21} = 2\left( P_{r,rb} + P_{r,rT} + P_{r,r\alpha} + P_{b,rb} + P_{b,rT} + P_{b,r\alpha} \right. \\
\left. + P_{T,rb} + P_{T,rT} + P_{T,r\alpha} + P_{\alpha,rb} + P_{\alpha,rT} + P_{\alpha,r\alpha}\right)\\
+P_{rb,rb} + P_{rT,rT} + P_{r\alpha,r\alpha} + 2P_{rb,rT} + 2P_{rb,r\alpha} + 2P_{rT,r\alpha}\\
+\frac{2}{3}\left(P_{m,rb} + P_{m,rT} + P_{m,r\alpha}\right).
\end{multline}
The full angle-averaged 21-cm power spectrum (linearised in $\delta_b,\,\delta_T,\,\delta_{\alpha}$) is then simply given by
\begin{equation}\label{Ptot}
P_{21} = P^{\rm (lin)}_{21} + P^{(nl)}_{21}.
\end{equation}
In the following sub-sections, we show how the individual components of Eqs.~(\ref{Plin}, \ref{Pnl}, \ref{Ptot}) can be obtained via the combination of flux, temperature, and matter profiles using the halo model approach.

\subsection{Halo model formalism}
The halo model consists of an analytical approach originally developed to compute the matter power spectrum at linear and nonlinear scales \citep{Seljak:2000gq,Peacock:2000qk,Cooray:2002dia}. It is based on the idea that our universe consists of haloes that are clustered around the peaks of linear matter perturbations. Despite these simplifying assumptions, the halo model has been shown to accurately predict the nonlinear matter power spectrum especially when empirical correction terms are included \cite[e.g.][]{Mead:2015yca,Murray:2020dcd}. 

However, the 21-cm signal is much more complex than the gravity-only clustering signal of the large-scale structure. It not only depends on halo properties but is also dominated by flux profiles that affect the gas of the IGM. Despite the very different physics at play, the formalism of the halo model has been shown to be well suited to the 21-cm signal at cosmic dawn \cite{Schneider:2020xmf}. Instead of halo profiles, the building blocks are radiation flux profiles that may overlap and extend over several hundreds of Mpc.

With the halo model formalism, the power spectra presented in Eq.~(\ref{Plin}) are obtained as follows:
\begin{eqnarray}\label{HM}
& P_{X,Y}^{1h}(k)  = & \frac{\beta_X\beta_Y}{\langle\rho_X\rangle\langle\rho_Y\rangle}\int dM \frac{dn}{dM} |W_X||W_Y|\,, \\
& P_{X,Y}^{2h}(k)  = & \frac{\beta_X}{\langle\rho_X\rangle}\int dM \frac{dn}{dM} |W_X|b_X \nonumber\\
& & \times \frac{\beta_Y}{\langle\rho_Y\rangle}\int dM \frac{dn}{dM} |W_Y|b_Y \times P_{\rm lin}\,,\\
&P_{X,Y}(k) = & P_{X,Y}^{1h}(k) + P_{X,Y}^{2h}(k)\,,
\end{eqnarray}
where $X$ and $Y$ correspond to the individual components $\lbrace r, b, T, \alpha, m \rbrace$. 
The halo mass function $dn/dM$ and bias $b_I(M)$ are calculated using the extended Press-Schechter (EPS) approach as described in Eq.~(1-4) of \citet{Schneider:2020xmf}. The EPS model comes with two parameters $(q,\,p)$ describing the shape of the first crossing distribution. The window functions $W_{I}$ represent the Fourier transforms of the profiles $\rho_{I}$, i.e.,
\begin{eqnarray}
W_{I}(k|z,M)=4\pi\int dr r^2 \rho_{I}(r|z,M)\frac{\sin(kr)}{kr}
\end{eqnarray}
with $I=\lbrace X,Y\rbrace$. The mean of the profile ($\langle\rho_I\rangle$) is given by
\begin{eqnarray}\label{rhoav}
\langle\rho_{I}\rangle=4\pi \int dM\frac{dn}{dM}\int dr r^2 \rho_{I}(z,M),
\end{eqnarray}
which corresponds to an average over the profile radii and the halo number density.

Before describing the shapes of the individual profiles ($\rho_I$) in more detail, let us introduce some important relations describing the sources and their photon emission. The star formation efficiency $f_*(M) \equiv \dot M_{*}/{\dot M}$ is given by
\begin{eqnarray}\label{fstar}
f_*(M) = \left(\frac{\Omega_b}{\Omega_m}\right)\frac{f_{*,0}\left[1+(M_t/M)^{\gamma_3}\right]^{\gamma_4}}{(M/M_p)^{\gamma_1} + (M/M_p)^{\gamma_2}}.
\end{eqnarray}
Note that as long as $f_*$ is assumed to be independent of redshift, it is equivalent to the stellar-to-halo mass ratio, i.e., $f_*(M)=M_{*}/M$.

The functional form of Eq.~(\ref{fstar}) is in agreement with expectations from galaxy abundance matching results at low and high-redshift \citep{Behroozi:2019kql}. It is based on the relation from Ref.~\citep{Mirocha:2017aaa} with an additional suppression (or boost) term at the truncation mass $M_t$ \citep[see][]{Schneider:2020xmf}. Reasonable default parameters are $M_p=2\times10^{11}$ M$_{\odot}$/h, $\gamma_1=-0.49$, and $\gamma_2=0.61$ \citep{Mirocha:2017aaa}. For the truncation parameters we assume $M_t=10^8$ M$_{\odot}$/h, $\gamma_3=1.4$, $\gamma_4=-4$.  

The escape fraction of ionising photons is given by
\begin{eqnarray}\label{fesc}
f_{\rm esc}(M) = f_{\rm esc,0}\left(M_{q}/M\right)^{\alpha_{\rm esc}},
\end{eqnarray}
with $M_q=10^{10}$ M$_{\odot}$/h following the parametrisation of Ref.~\cite{Park:2018ljd}. The function is truncated at $f_{\rm esc}\leq 1$. A positive power-law index $\alpha_{\rm esc}$ ensures that smaller galaxies with lower gas column densities have higher escape fractions in agreement with simulations \citep[see e.g. Ref.][]{Paardekooper:2015via}. A reasonable default setting is $f_{\rm esc,0}=0.1$ and $\alpha_{\rm esc}=0.5$ \cite{Park:2018ljd}.

The spectral properties of the sources are parametrised via the emissivity parameters ($\varepsilon_I$). They are defined for three broad bands, the Lyman-$\alpha$ radiation band causing the absorption signal via to the Wouthuysen-Field effect \cite{Wouthuysen1952aaa,Field1958aaa}, the X-ray band leading to the heating of the gas, and the UV radiation band inducing the reionization process. The emissivity of the Lyman-$\alpha$ and the UV bands are parametrised as
\begin{equation}
\varepsilon_{i}(\nu)=\frac{N_{i}}{m_p}I_{i}(\nu)
\end{equation}
where $m_p$ is the proton mass and $N_{i}$ the number of photons in the given band $i=\lbrace\alpha,r \rbrace$. The spectral properties are parametrised in $I_{i}(\nu)$ as shown below. For the X-ray band, we define
\begin{equation}
\varepsilon_{X}(\nu)=\frac{f_X c_X}{\nu h_P}I_{i}(\nu),
\end{equation}
where $i=X$. The free parameter $f_X$ has a default value of one and $c_X=3.4\times10^{40}$ erg yr s$^{-1}$ M$_\odot^{-1}$ is constrained via low-redshift observations \cite{Fragos:2012vf,Das:2017fys}.

For the spectral properties of each band, we assume a simple power-law behaviour, i.e.,
\begin{equation}
I_i(\nu) = A_i\nu^{\alpha_i} 
\end{equation}
where $A_i$ is a normalisation factor which is set so that the integral (between $\nu_{\rm i, min}$ and $\nu_{\rm i, max}$) over $I_i(\nu)$ becomes unity. Note that we have assumed $i=\lbrace\alpha,r,X\rbrace$ here.

Now that we have established the parametrisation of the source properties, let us have a closer look at the flux profiles. We only proved an overview here, for more details we again refer to Ref.~\citep{Schneider:2020xmf}.

\begin{figure*} 
\centering
\includegraphics[width=0.95\textwidth,trim=0.2cm 0.1cm 0.1cm 0.1cm, clip]{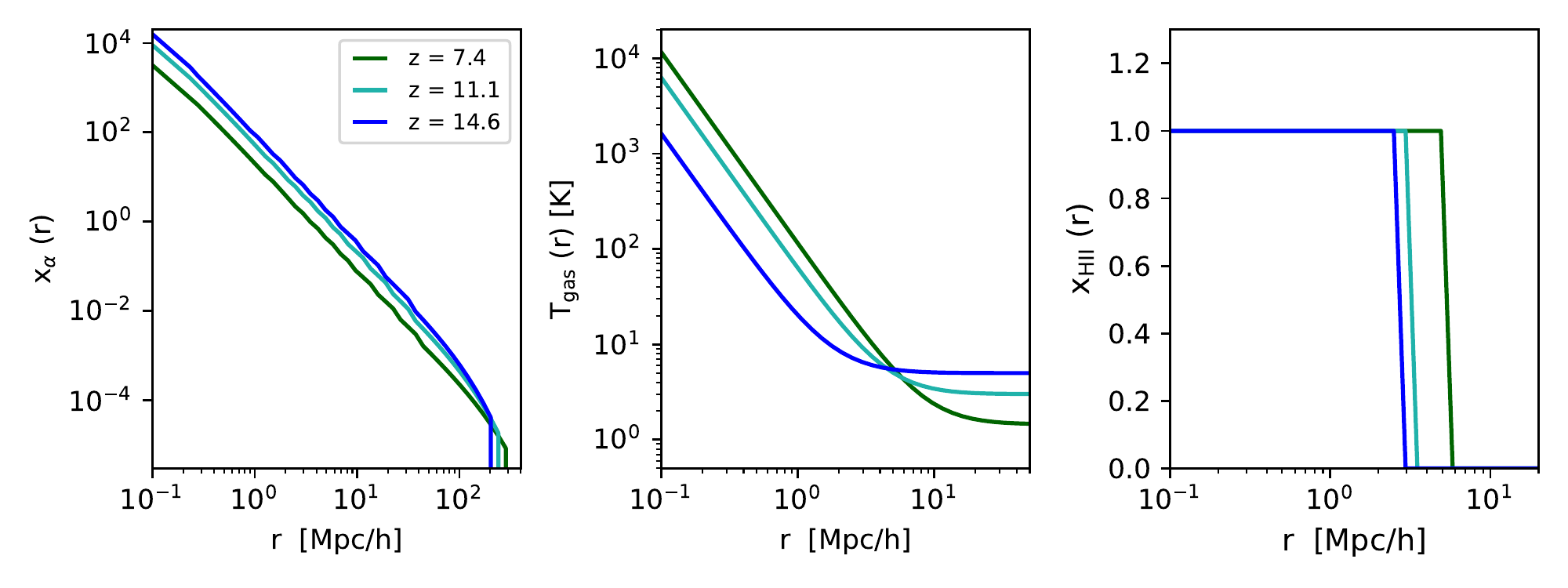}
\caption{\label{fig:profiles}Example of the radiation coupling coefficient $x_{\alpha}$ (left), the gas temperature $T_{\rm gas}$ (middle), and the reionization profile $x_{\rm HII}$ (right) for a halo with fixed mass $M=10^{11}$ M$_{\odot}$/h at three different redshifts.}
\label{fig:profiles}
\end{figure*}

\subsubsection{Matter and baryonic profiles}
The density profile of the total matter component ($\rho_m$) is modelled assuming a NFW profile \cite{Navarro:1995iw} with concentration parameters depending on cosmology \citep[following the approach of Ref.][]{Schneider:2021wds}. This corresponds to the approach used in the standard halo model of structure formation. For the baryonic profile, we assume $\rho_b= (\Omega_b/\Omega_m)\rho_m$. While the true profile inside haloes is likely deviating from the NFW shape due to cooling and feedback effects \cite[e.g.][]{giri2021emulation}, this is unlikely to affect the scales we are interested in. Note, furthermore, that, although the baryonic profile is suppressed in absolute amplitude compared to $\rho_m$, this has no influence on the perturbations, i.e., $\delta_b=\delta_m$.

\subsubsection{Lyman-$\alpha$ flux profiles}
The radiation of the first stars in the range between the Lyman-$\alpha$ and the Lyman limit frequencies causes a spin-flip of the hydrogen atoms in the IGM, forcing the spin temperature out of equilibrium with the background temperature of the CMB \cite{Wouthuysen1952aaa,Field1958aaa}. This effect leads to an absorption feature in the CMB spectrum at radio frequencies during the epoch of cosmic dawn, i.e. between $z\sim10-25$. The flux profile of the Lyman-$\alpha$ radiation is given by
\begin{multline}\label{rholyal}
\rho_{\alpha}(r|M,z) = \frac{(1+z)^2}{4\pi r^2}\sum_{n=2}^{n_m}f_n\varepsilon_{\alpha}(\nu') f_* {\dot M}_{\rm ac}(z'|M,z),
\end{multline}
where $\nu'=\nu (1+z')/(1+z)$ and where $f_n$ are the recycling fractions from Ref.~\citep{Pritchard:2005an} assuming the truncation $n_m=23$. Eq.~(\ref{rholyal}) contains the $1/r^2$ volume effect and depends on the mass accretion ($\dot{M}_{\rm ac}$) of the corresponding source. Different methods to calculate $M_{\rm ac}(z)$ are discussed in \citet{Schneider:2020xmf}. The profile furthermore depends on look-back redshift $z'$, which corresponds to the redshift when a given photon at radius $r$ was emitted by the source. The look-back redshift is obtained by inverting the comoving distance relation
\begin{equation}
r(z'|z) = \int_z^{z'}\frac{c}{H(z'')}dz'' 
\end{equation}
where $c$ is the speed of light and $H(z)$ the Hubble parameter.

The Lyman-$\alpha$ flux profile is connected to the radiation coupling coefficient via the relation
\begin{equation}
x_{\alpha}(r|M,z) = \frac{1.81\times10^{11}}{(1+z)}S_{\alpha}(z)\rho_{\alpha}(r|M,z),
\end{equation}
where the suppression factor $S_{\alpha}(z)$ is obtained from Ref. \citep{Furlanetto:2006jb}.

In the left-hand panel of Fig.~\ref{fig:profiles}, we plot three typical $x_{\alpha}$ profiles from sources at different redshifts each source residing in a halo of $M=10^{11}$ M$_{\odot}$/h (coloured lines). The profiles are characterised by a $1/r^2$ slope that becomes gradually steeper towards larger radii due to the look-back effect. They furthermore exhibit the typical step-like features as shown in \cite{Holzbauer:2011mv,Schneider:2020xmf}.

\subsubsection{Temperature profiles}
The 21-cm signal crucially depends on the temperature of the IGM. From Eq.~(\ref{dTb}) it is obvious that once the gas is heated up beyond the background CMB temperature, the signal changes sign, switching from absorption to emission. The temperature of the IGM is predominantly regulated by the interactions between hydrogen atoms and the radiation from the first sources, the most efficient photons originating from the X-ray regime.

The flux profile of the X-ray photons is given by
\begin{multline}\label{rhoxray}
\rho_{\rm xray}(r|M,z)= \frac{(1+z)^2}{4\pi r^2}\sum_{i}f_if_{X,h}\\
\times\int_{\nu_{\rm th}^i}^{\infty} d\nu (\nu-\nu_{\rm th}^i)h_P\sigma_i(\nu)\varepsilon_{X}(\nu'){\rm e}^{\tau_{\nu'}}f_* {\dot M}_{\rm ac}(z'|M,z),
\end{multline}
where $i=\lbrace {\rm H,He}\rbrace$ and $\nu_{\rm th}^i$ and $\nu_{\rm th}^{i}h_P=\lbrace13.6,26.5\rbrace$ eV. More details regarding the cross sections ($\sigma_i$) and the optical depth ($\tau_\nu$) can be found in \citet{Schneider:2020xmf}. Note that Eq.~(\ref{rhoxray}) is provided in physical units.

From the X-ray flux profiles, we calculate the heating profiles ($\rho_h$) via the equation
\begin{equation}
\frac{3}{2}\frac{d\rho_h(r|M,z)}{dz}=\frac{3\rho_h(r|M,z)}{(1+z)}-\frac{\rho_{\rm xray}(r|M,z)}{k_B(1+z)H(z)},
\end{equation}
where $k_B$ is the Boltzmann constant. 

Some examples of temperature profiles at different redshifts are shown in the middle panel of Fig.~\ref{fig:profiles}. Instead of plotting the heating profiles directly, we show the total gas temperature $T_{\rm gas}(r)=\rho_h(r)+T_{\rm ad}$, where $T_{\rm ad}\propto (1+z)^{-2}$. The gas temperature roughly follows a $1/r^2$ power law before transitioning towards a constant value given by the adiabatic background temperature ($T_{\rm ad}$) of the IGM.


\subsubsection{Reionization profiles}
The reionization bubbles are caused by photons with energies above 13.6 eV that ionize the IGM. Due to the large optical depth in that regime of the spectrum, the bubbles are assumed to have well-defined borders. We can therefore approximate the reionization bubbles by a profile of the form
\begin{equation}
\rho_{\rm reio}(r|M,z)= \theta_H\left[R_{b}(M,z)-r\right],
\end{equation}
where $\theta_H$ is the Heaviside step function and $R_{b}$ is the bubble radius. The latter is not straightforward to calculate. We use the excursion-set method introduced by \citet{Furlanetto:2004nh} to obtain a distribution of bubbles as a function of their size $R_{b}$. The method is summarised in Sec.~\ref{sec:BubbleExc}. For the sake of the reionization profile, it is important to notice that most bubbles contain multiple sources. It is therefore unclear how to connect the bubble radius to the underlying halo (with mass $M$). We solve this problem by applying an abundance matching method, which matches the largest bubbles to the largest haloes in a strictly hierarchical sense.

In the right-hand panel of Fig.~\ref{fig:profiles}, we illustrate three examples of ionising bubbles at different redshifts. Each bubble is abundance matched to a halo of $M=10^{11}$ M$_{\odot}$/h at that given redshift. Hence, the plot should not be confused with a growing profile from one individual expanding bubble. 

\subsection{Bubble-halo connection} \label{sec:BubbleExc}
In order to include bubbles into the halo model formalism, we abundance-match bubbles to haloes, which means we need to know the number of bubbles as a function of their radii. The excursion-set formalism first introduced in Ref.~\cite{Furlanetto:2004nh} provides a convenient way to obtain this information. It is based on the barrier
\begin{equation}
B(S_0)= \frac{F_{(q,p)} N_{\rm ion}}{n_{\rm rec}} \int dM f_{\rm esc}f_{*} \frac{dn(\delta,S|\delta_0,S_0)}{dM}
\end{equation}
where $S(M)$ is the variance of the matter field. The specific variance $S_0=S(M_0)$ corresponds to a smoothing scale $M_0$ with corresponding density field $\delta_0$. The conditional mass function is given by
\begin{equation}\label{condMF}
\frac{dn(\delta,S|\delta_0,S_0)}{dM}= \frac{(\delta-\delta_0)}{\sqrt{2 \pi(S-S_0)^3}}{\rm e}^{\frac{(\delta-\delta_0)^2}{2(S-S_0)}} \frac{dS}{dM}.
\end{equation}
Note that Eq.~(\ref{condMF}) is based on the \citet{Press:1973iz} description of spherical collapse ($q=1$, $p=0$) while the halo mass function use above is based on a more general prescription (where $q$ and $p$ are free model parameters). In  order to correct for this, we follow Ref.~\cite{Mesinger:2011aaa} and introduce the correction factor
\begin{equation}
F_{(q,p)} = \left. \int dM f_{\rm esc}f_{*}\frac{dn_{(q,p)}}{dM} \middle/ \int dM f_{\rm esc}f_{*}\frac{dn_{(1,0)}}{dM}\right.
\end{equation}
with $n_{(1,0)}$ and $n_{(q,p)}$ being the Press-Schechter \cite{Press:1973iz} and the generalised halo mass function, respectively.

\begin{figure*}  
\centering
\includegraphics[width=0.49\textwidth,trim=0.3cm 0.1cm 1.2cm 0.5cm, clip]{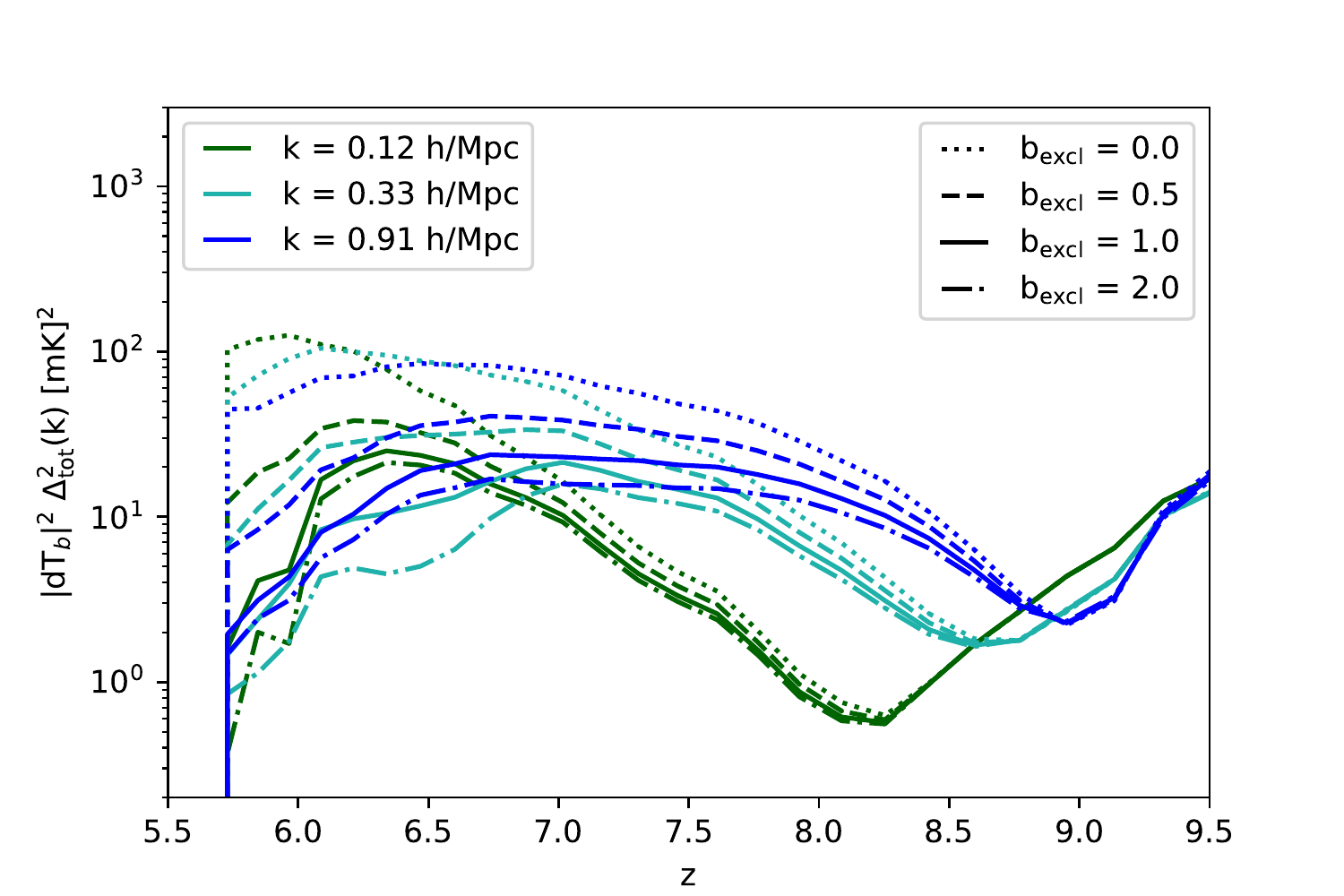}
\includegraphics[width=0.49\textwidth,trim=0.3cm 0.1cm 1.2cm 0.5cm, clip]{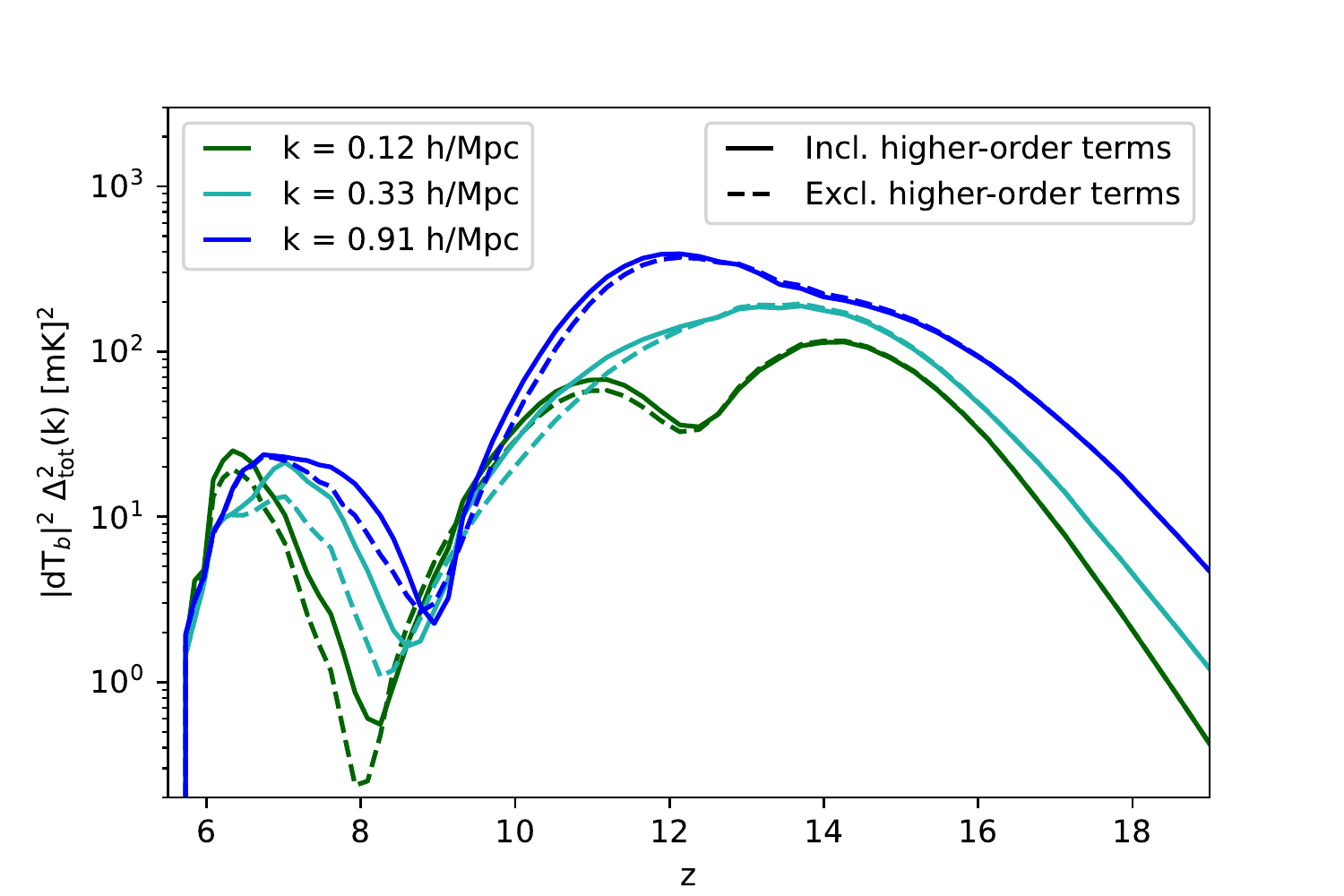}\\
\caption{Effects of the empirical bubble-overlap correction factor (left) and the higher-order terms (right) on the 21-cm power spectrum. Different colours correspond to different $k$-modes.}
\label{fig:overlap_HO}
\end{figure*}

Following \cite{Furlanetto:2004nh}, we linearise the barrier $B_{\rm lin}=B_0 + \beta_0 S$, where $B_0=B(0)$ and $\beta_{0} = dB/dS|_{S=0}$. Since the excursion set calculation becomes analytical for a linear barrier, we can write
\begin{equation}\label{bubblefct}
\frac{dn_{b}}{d\ln M_b} = \frac{{\bar\rho}}{M_b}\frac{B_0}{\sqrt{2\pi S}}\exp\left[-\frac{B_{\rm lin}^2}{S}\right]\frac{d\ln S}{d\ln M_b}
\end{equation}
to obtain the number density of reionization bubbles as a function of their size $R$. Note that Eq.~\ref{bubblefct} does not directly depend on halo mass or sources but is a function of the average mass ($M_b$) enclosed by the bubble radius ($R_b$). It is therefore unclear which sources correspond to which bubbles. We solve this problem by abundance matching haloes to bubbles in a strictly hierarchical way. In practice this means we start with the largest bubbles matching them to the largest haloes before going to smaller and smaller bubbles until they are all matched to haloes.

Another open question is how to quantify the bias of reionization bubbles. One possibility is to attribute to each bubble the bias of the matched halo. However, as bubbles are evolving around several sources, such a halo bias is likely too high. On the other hand, the excursion set algorithm cannot only be used to obtain a bubble mass function but also a bubble bias prescription. Following \citet{McQuinn:2005ce} we use
\begin{equation}\label{bubblebias}
b_r(M) = 1 + \frac{1}{D(z)}\left[\frac{B_{\rm lin}}{S} - \frac{1}{B_0}\right],
\end{equation}
where $D(z)$ is the growth factor. Note that while Eq.~(\ref{bubblebias}) is positive for most cases, it can become negative if the corresponding bubbles are sufficiently small. Since a negative bubble bias leads to unwanted features in Eq.~(\ref{HM}), we set all negative values of Eq.~(\ref{bubblebias}) to zero.

\subsection{The issue of bubble overlap}
The excursion-set approach for reionised bubbles \citep{Furlanetto:2004nh} evades the problem of bubble overlap in the sense that it creates bubbles from overdensity regions containing many sources per bubble. However, there remains the problem that we assume spherically symmetrical bubbles while in reality bubbles will have complex shapes, especially towards the end of reionization. As it is impossible to fill a volume with packed spheres, the halo model approach is inherently incapable to model the very late times of reionization when the power spectrum signal gradually disappears and only a few neutral islands remain.

A further problem with the formalism of the halo model arises due to the fact that Eq.~(\ref{HM}) explicitly allows for the overlap of bubbles despite the fact that they are already included in the excursion set calculation. A potential way to deal with this issue is to add a bubble exclusion term to the halo model formalism. Such an approach has been proposed in the context of the standard halo model of large-scale structure \citep{Smith:2010fh,vandenBosch:2012nq} where haloes are not allowed to overlap either.

The exclusion term derived in \citep{Smith:2010fh} is negative and has a shape very similar to the one-halo power spectrum. For simplicity, we therefore abstain from performing the full calculation but simply suppress the one-halo term by imposing
\begin{equation}
P_{r,r}^{(\rm 1h)}\rightarrow c_{\rm excl}^2(z)P_{r,r}^{(\rm 1h)},\hspace{0.5cm}P_{r,X}^{(\rm 1h)}\rightarrow c_{\rm excl}(z)P_{r,X}^{(\rm 1h)}
\end{equation}
with $X=\lbrace b,T,\alpha,m\rbrace$. The empirical correction factor $c_{\rm excl}(r)$ is given by
\begin{equation}
c_{\rm excl}=(1-{\langle\rho}_{\rm reio}\rangle)^{b_{\rm excl}},
\end{equation}
where $b_{\rm excl}$ is a free model parameter that can be fixed using simulations. It consists of the only truly empirical parameter in the model presented here. 

The effect of the bubble overlap parameter $b_{\rm excl}$ is illustrated in the left-hand panel of Fig.~\ref{fig:overlap_HO}. The panel shows the redshift dependence of the power spectrum (for three different $k$-modes) during the reionization epoch. Different linestyles correspond to different values of $b_{\rm excl}$, the higher the value the stronger the suppression of the power spectrum at late times. The best match with results from {\tt 21cmFAST} is found for values of $b_{\rm excl}=1-1.5$ (see Sec.~\ref{sec:comparison}).

\subsection{Higher order terms}
After having discussed how to calculate the individual auto and cross spectra from Eq.~(\ref{Plin}), let us turn to the higher-order terms introduced in Eq.~(\ref{Pnl}). Since the bubble ionisation field may be nonlinear even at small $k$-modes, we have to include these terms in the analysis \cite[see e.g. Refs.][]{Lidz:2006vj,Georgiev:2021yvq}.

In principle, it would possible to obtain higher-order terms with the halo model by multiplying individual profiles before using the general formalism of Eq.~(\ref{HM}). However, we prefer a much simpler, approximate calculation based on the lower-order power spectra, i.e.,
\begin{eqnarray}
&P_{r,rX}&= P_{r,r} \left(\frac{k^3 P_{X,X}}{2\pi^2}\right)^{\frac{1}{2}},\nonumber\\
&P_{X,rY}&= P_{r,X} \left(\frac{k^3 P_{Y,Y}}{2\pi^2}\right)^{\frac{1}{2}},\\
&P_{rX,rY}&= P_{r,X} \left(\frac{k^3 P_{r,Y}}{2\pi^2}\right),\nonumber
\end{eqnarray}
where $X,Y=\lbrace b,T,\alpha\rbrace$. The above equations are used to obtain all terms from Eq.~(\ref{Pnl}). While they are not expected to provide precise results, we will see in Sec.~\ref{sec:comparison} that the approach is good enough for our current needs.

In the right-hand panel of Fig.~\ref{fig:overlap_HO} we plot the 21 cm power spectrum as a function of redshift for both cases when higher-order terms are included (solid lines) or left out (dashed lines). In general, the higher-order terms lead to a moderate increase of the temperature and the reionization peaks of the power spectrum. The effect is present for both small and large $k$-modes. The changes are typically at the level of about 20 percent but can grow as large as a factor of two especially in the trough between the temperature and ionisation peaks.

\subsection{The global signal}
The global differential brightness temperature corresponds to the 21-cm signal averaged over the entire sky. In terms of the halo model, average quantities are obtained by integrating the profiles of the individual components over the volume and the source abundance as defined in Eq.~(\ref{rhoav}). From Eq.~(\ref{dTb}) it is then straightforward to obtain the global signal
\begin{equation}\label{GS}
\langle dT_b\rangle =  T_0 \langle x_{\rm HI}\rangle \frac{\langle x_{\alpha}\rangle}{1+\langle x_{\alpha}\rangle}\left[1-\frac{T_{\rm cmb}}{\langle \rho_h\rangle + T_{\rm ad}}\right],
\end{equation}
with $\langle x_{\rm HI}\rangle=1- \langle\rho_{\rm reio}\rangle$. Note that with the calculation presented above, we now use the same method for the global signal and the power spectrum. This was not the case in the original model presented in \citet{Schneider:2020xmf} where we relied on an approach inspired by Ref.~\cite{Mirocha:2014faa}.

\begin{figure*} 
\centering
\includegraphics[width=0.99\textwidth,trim=0.7cm 0.1cm 1.6cm 0.5cm, clip]{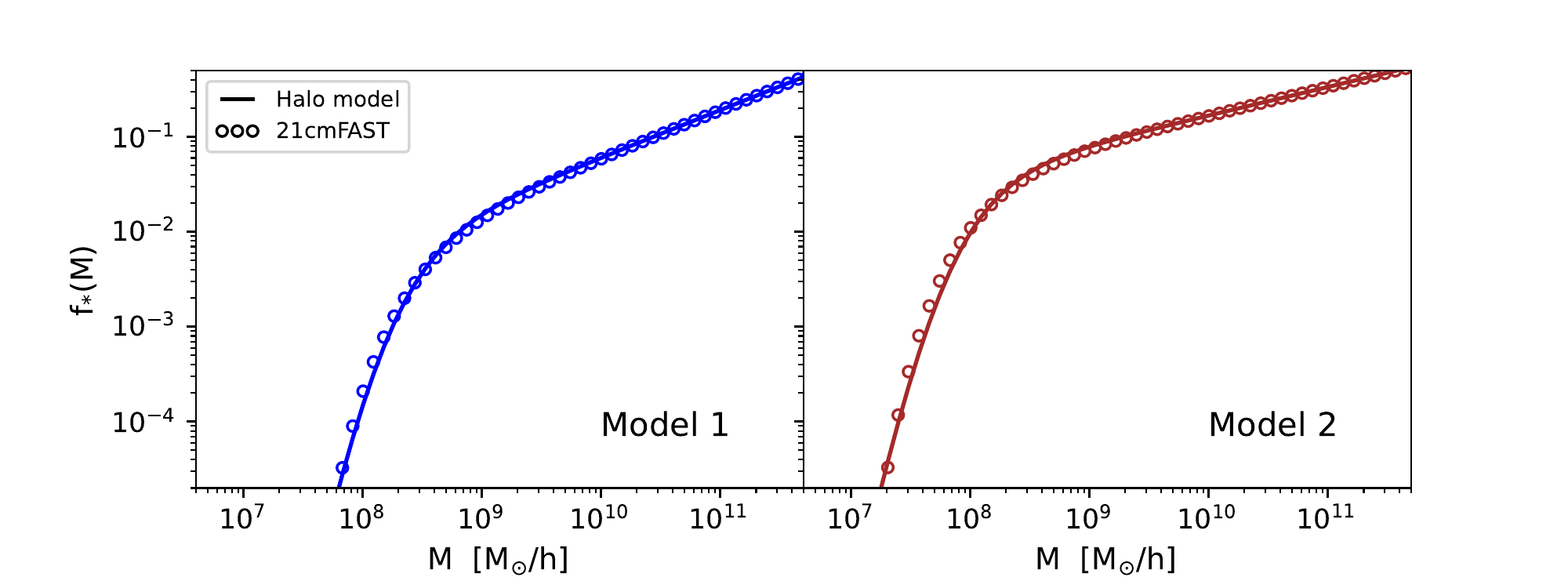}
\caption{Stellar-to-halo mass ratio for Model 1 (left) and Model 2 (right) of {\tt 21cmFAST} (empty circles). The matched functions used in {\tt HMreio} (corresponding to Eq.~\ref{fstar} with parameters from \ref{fstarparam1} and \ref{fstarparam2}) are shown as solid lines.}
\label{fig:comparison1}
\end{figure*}

\section{Comparison to 21cmFAST}\label{sec:comparison}
In this section, we validate the halo model of reionization ({\tt HMreio}) by comparing it to results from the semi-numerical code {\tt 21cmFAST} \citep{Mesinger:2011aaa,Murray2020aaa}. Note that this comparison is not straightforward as there are several subtle differences in the setup and the parametrisation between the two codes. For example, {\tt 21cmFAST} uses a star-formation rate that is proportional to the stellar mass times the Hubble parameter, while we rely on an excursion-set model instead \footnote{Note that there is also the option to assume an exponential growth or an abundance-matching approach, see \citet{Schneider:2020xmf}}. Another example is the parametrisation of $f_{*}(M)$ which, in the case of {\rm 21cmFAST}, features an exponential truncation at small scales, whereas we apply the parametrisation presented in Eq.~(\ref{fstar}). Further differences related to the number and distribution of sources, the spectral property of the Lyman-$\alpha$ radiation, or the calculation of the gas temperature are likely to affect the results as well. This means that a strictly one-to-one comparison becomes impossible.

In order to carry out a meaningful comparison between the codes, we proceed in the following way: (i) we fit our model of $f_{*}(M)$ to the one from {\tt 21cmFAST}; (ii) we compare the mean ionisation fractions calculated with the two codes and calibrate the number of ionising photons ($N_{\rm ion}$) to obtain an optimal match; (iii) we investigate the absorption signal of the mean differential brightness temperature re-calibrating, if necessary, the total Lyman-$\alpha$ and X-ray radiation fluxes ($N_{\alpha}$, $f_X$) in order to obtain a better agreement; (iv) we compare the power spectrum as a function of redshift for different $k$-modes between 0.1 and 1 h/Mpc.

Before calibrating the model, let us specify the main modelling and parameter choices of {\tt HMreio} that we use for the comparison and the subsequent forecast analysis. For the first-crossing distribution of the halo mass function and the bias we use $q=0.85$ and $p=0.3$ which yields results between the \citet{Press:1973iz} and the \citet{Sheth:2001dp} models and has been shown to agree with simulations at $z\sim 5-20$ \cite{Schneider:2018xba}. The mass-accretion rate is modelled using the EPS model presented in \citet{Schneider:2020xmf}. For the spectral energy range of the X-ray radiation, we assume $E_{\rm min}=500$ and $E_{\rm max}=2000$. In between this range, the spectrum is assumed to decrease as a power law with $\alpha_X=1$. The spectra from the Lyman-$\alpha$ and the ionising radiation are assumed to be flat ($\alpha_{\alpha}=\alpha_{r}=0$). All these assumptions agree with the parametrisation used in {\tt 21cmFAST}.

We run two different {\tt 21cmFAST} simulations for the comparison. The first simulation (Model 1) corresponds to the benchmark run from \citet{Park:2018ljd}. It is based on the parameters $N_{\gamma}=5000$, $f_{*,10}=0.05$, $\alpha_*=0.5$, $f_{\rm esc,0}=0.1$, $\alpha_{\rm esc}=-0.5$, $M_{\rm turn}=5\times10^{8}$ M$_{\odot}$, and $\log_{10}(L_X/{\rm SFR}) = 40.5$ providing a good match to the luminosity functions from the Hubble Space Telescope. For the second simulation (Model 2) we assume $N_{\gamma}=5000$, $f_{*,10}=0.15$, $\alpha_*=0.3$, $f_{\rm esc,0}=0.03$, $\alpha_{\rm esc}=0.0$, $M_{\rm turn}=2\times10^{8}$ M$_{\odot}$, and $\log_{10}(L_X/{\rm SFR}) = 40.2$, instead, which corresponds to a model with reduced X-ray radiation, constant escape fraction, and larger stellar-to-halo mass ratio extending to smaller halo masses. As a consequence, Model 2 yields a more prominent global absorption signal starting at a higher redshift as well as a longer period where the power spectrum is dominated by the reionization process. We refer to \citet{Mesinger:2011aaa} and \citet{Park:2018ljd} for the definitions of the {\tt 21cmFAST} parameters mentioned above.

As discussed above, we first fit our model to the stellar-to-halo mass ratio of {\tt 21cmFAST}. The result of this fitting process is shown in Fig.~\ref{fig:comparison1}, where empty circles and solid lines correspond to $f_*(M)$ from {\tt 21cmFAST} and {\tt HMreio}, respectively. In terms of the {\tt HMreio} model parameters, the best-fitting values are
\begin{equation}\label{fstarparam1}
f_{*,0}=0.05,\,\, M_t=2.0\times10^8\,\, {\rm M_{\odot}},\,\,\gamma_1=\gamma_2=-0.5,
\end{equation}
for Model 1 and
\begin{equation}\label{fstarparam2}
f_{*,0}=0.15,\,\, M_t=8.2\times10^7\,\, {\rm M_{\odot}},\,\,\gamma_1=\gamma_2=-0.3,
\end{equation}
for Model 2. The remaining parameters $M_p=10^{10}$, $\gamma_3=1.4$, and $\gamma_4=-4$ are the same for both models.

\begin{figure*} 
\centering
\includegraphics[width=0.99\textwidth,trim=0.7cm 0.1cm 1.6cm 0.5cm, clip]{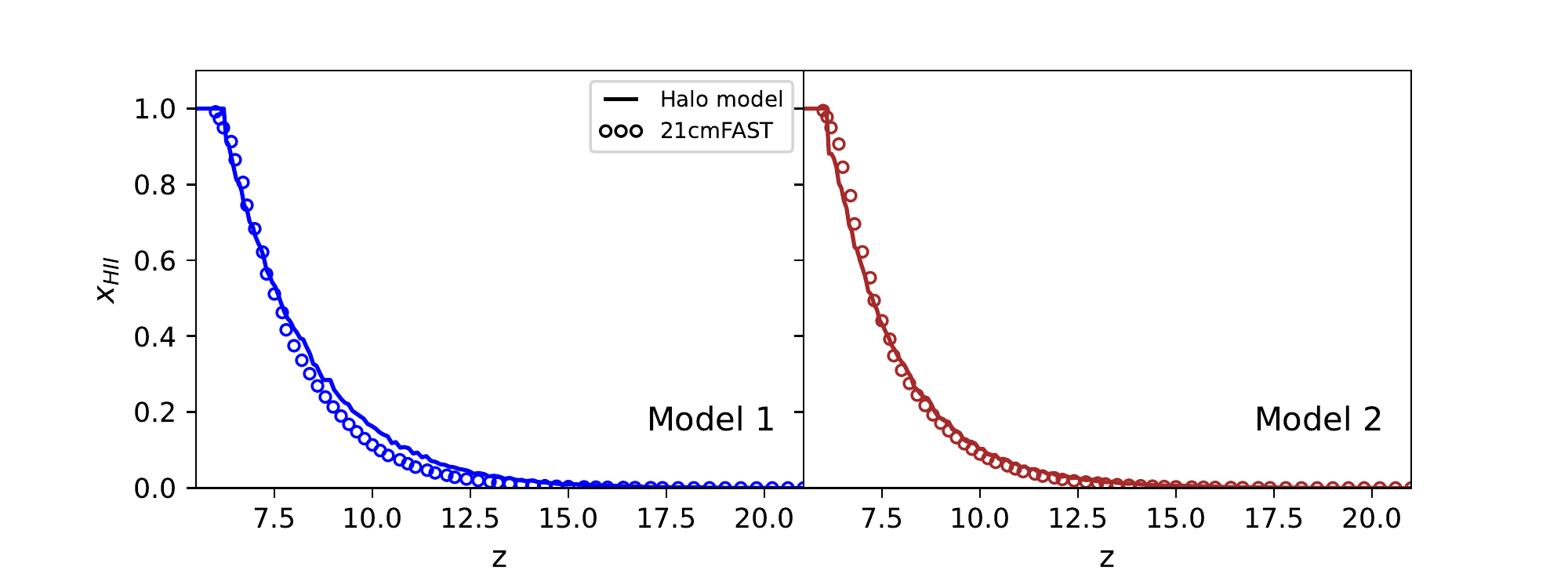}\\
\includegraphics[width=0.99\textwidth,trim=0.7cm 0.1cm 1.6cm 0.5cm, clip]{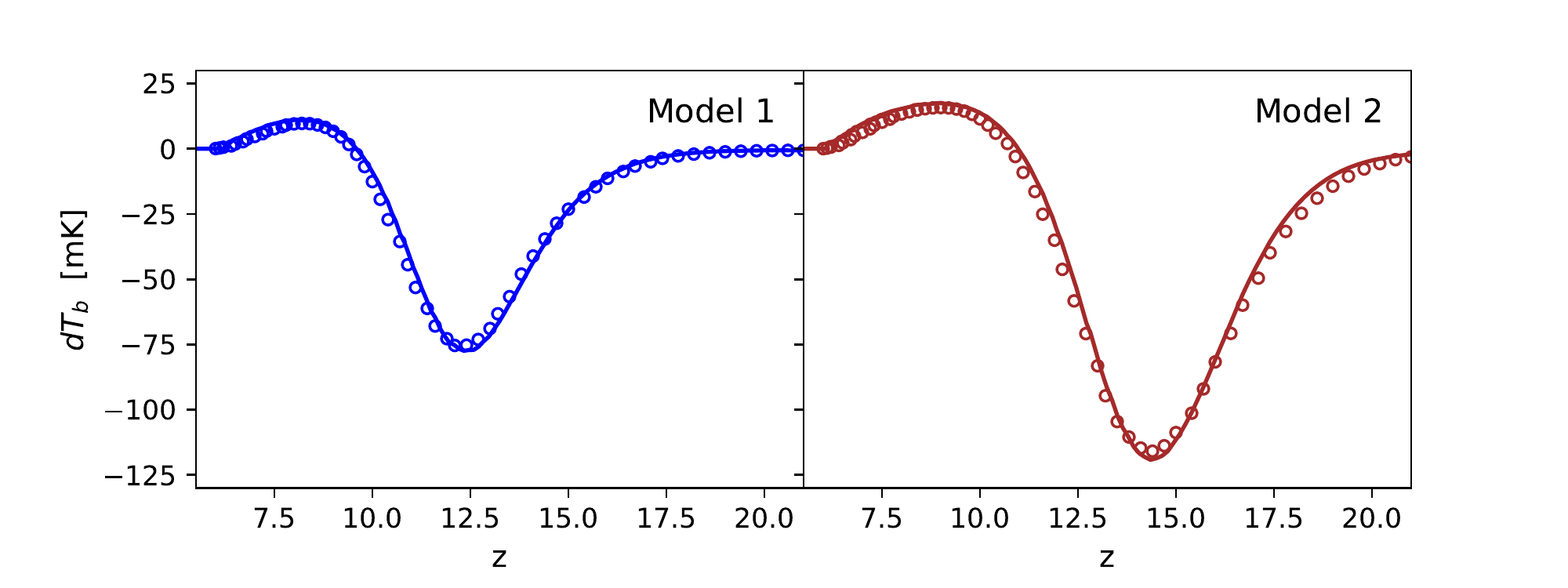}\\
\includegraphics[width=0.99\textwidth,trim=0.7cm 0.1cm 1.6cm 0.5cm, clip]{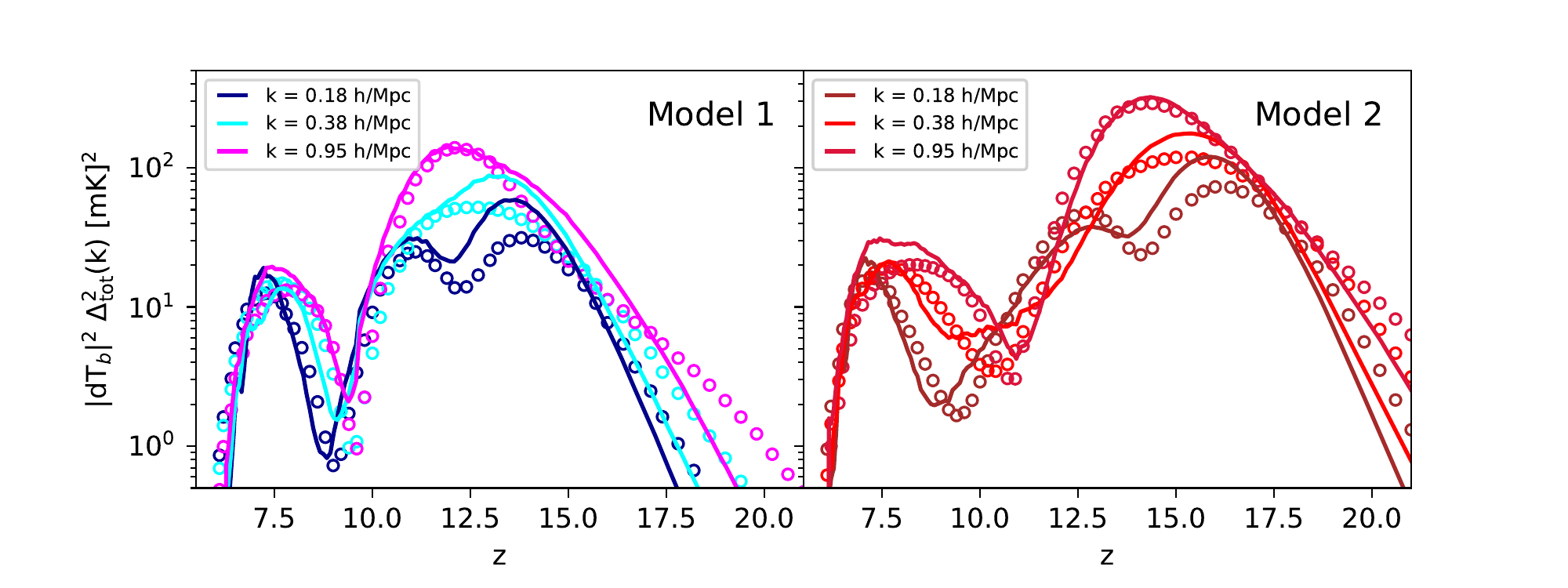}
\caption{Comparison of {\tt 21cmFAST} (empty circles) with the halo model of reionization (solid lines). From top to bottom we illustrate the global reionization fraction, the global differential brightness temperature, and the power spectrum as a function of redshift. \emph{Left:} simulation with default parameters (Model 1). \emph{Right:} simulation with increased star formation, lower escape fraction, and reduced X-ray radiation (Model 2).}
\label{fig:comparison2}
\end{figure*}

As a next step, we normalise the ionising photon number $N_{\rm ion}$ by comparing the global ionisation rate to the one from {\tt 21cmFAST}. A reasonably good agreement is found when using $N_{\rm ion}=6000$ for both Model 1 and 2. A comparison of the global ionisation fraction is shown in the top panels of Fig.~\ref{fig:comparison2}, where the results from {\tt 21cmFAST} and {\tt HMreio} are again shown as empty circles and solid lines. Note that the ionising photon number ($N_{\rm ion}$) is 20 percent higher than the one assumed in {\tt 21cmFAST} ($N_{\gamma}$). This is a direct consequence of the differences in the modelling of the number of sources between the two codes.

After specifying the ionizing photon number ($N_{\rm ion}$), we fit the Lyman-$\alpha$ ($N_{\alpha}$) and X-ray ($f_X$) fluxes based on the global differential brightness temperature.  For the Lyman-$\alpha$ flux, we find a good match assuming $N_{\alpha}=3500$ for both models. Note that there is no analogous number in {\tt 21cmFAST}, where the Lyman-$\alpha$ flux is described by a composite spectral model instead of a simple power law. Regarding the X-ray flux, the best agreement is obtained by assuming $f_X=0.9$ for Model 1 and $f_X=0.45$ for Model 2. When multiplied with our default value of $c_X=3.16\times 10^{40}$ $\rm erg\,yr\,s^{-1}M_{\odot}^{-1}$, this leads to a number that is about 20 percent smaller than the assumed value for $L_X$ in {\tt 21cmFAST}.

In the middle panels of Fig.~\ref{fig:comparison2} we show the global differential brightness signal from {\tt 21cmFAST} (empty circles) and {\tt HMreio} (solid lines) using the parameter values specified above. The global absorption signal of Model 2 is significantly deeper and shifted to higher redshifts compared to the one from Model 1. This is due to the increased stellar-to-halo fraction and, to a lesser degree, the reduced X-ray radiation. Note, furthermore, that although the signal is shifted to earlier times, the reionization process ends at approximately the same redshift, which is a consequence of the reduced $f_{\rm esc}$ of Model 2 counteracting the larger star formation efficiency.

Comparing the global signal from {\tt HMreio} and {\tt 21cmFAST} shown in Fig.~\ref{fig:comparison2}, we notice that the predictions are similar but not identical. The similarity is not very surprising, as we have been calibrating the parameters $N_{\alpha}$ and $f_X$ based on this plot. The remaining small differences, on the other hand, indicate that different modelling choices regarding e.g. the mass accretion and spectral properties of sources cannot be completely eradicated by re-calibrating the free model parameters.

In the bottom panels of Fig.~\ref{fig:comparison2} we compare the power spectra from {\tt 21cmFAST} (empty circles) and HMreio (lines) as a function of redshift for three different $k$-modes between 0.1 and 1 h/Mpc. Note that this plot has not been used during the calibration process. It therefore provides a first true test for the halo model of reionization. In general, we obtain a good qualitative agreement with {\tt 21cmFAST}. At large scales ($k=0.18$ h/Mpc), both codes exhibit the characteristic three peaks related to the Lyman-$\alpha$, the heating, and the reionization epochs. At smaller scales ($k=0.38$ and $0.95$ h/Mpc), the first two peaks merge to one in agreement with numerous previous results \cite[e.g.][]{Pritchard:2006sq,Pritchard:2008da}.

At the quantitative level, there are visible differences between the power spectra from {\tt HMreio} and {\tt 21cmFAST}. Over most redshifts, they remain at the level of a factor of two or less. The largest discrepancies appear at high redshifts around the Lyman-$\alpha$ peak which tends to be more pronounced in the case of {\tt HMreio} compared to {\tt 21cmFAST}. We speculate that these differences could be caused by different biasing assumptions of the very high-redshift sources or it could point towards differences in the treatment of shot noise \citep{Reis:2021sqh}. Another notable discrepancy can be observed at $z\sim7.5$ where the results from {\tt HMreio} tend to exceed the ones from {\tt 21cmFAST}. As this corresponds to the regime of bubble coalescence, we suspect that the model for bubble overlap in {\tt HMreio} might not be precise enough in this regime. Further investigations in the future are needed to clarify these points.


\begin{figure*} 
\centering
\includegraphics[width=0.98\textwidth,trim=0.6cm 0.1cm 1.5cm 0.5cm, clip]{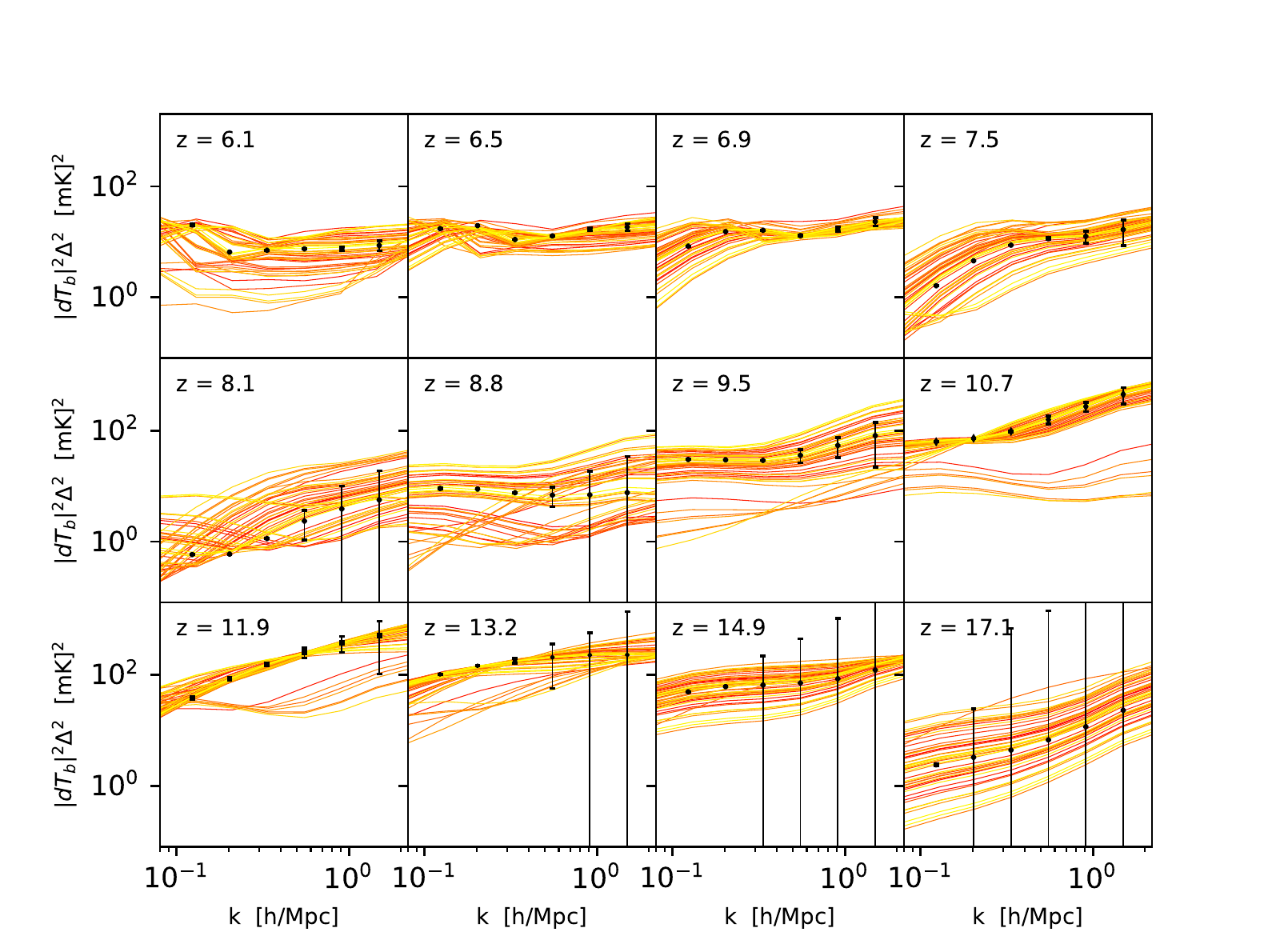}\\
\caption{Mock data of the 21 cm power spectrum for 1000h observations with SKA low assuming instrumental noise sample variance and scale cuts due to foreground contamination (black data points). The coloured lines correspond to 50 power spectra predicted by {\tt HMreio} with fixed astrophysical parameters and cosmological parameter values randomly drawn from a multivariate Gaussian posterior distribution  from {\tt Planck}.}
\label{fig:SKAmock}
\end{figure*}

\section{Forecast for SKA}\label{sec:forecast}
The main advantage of the halo model of reionization compared to other methods is that it can predict the 21-cm global signal and power spectrum within seconds. As a consequence, it becomes possible to efficiently scan the vast space of astrophysical and cosmological parameters. As a first application of the model, we perform a Monte-Carlo Markov-Chain (MCMC) forecast analysis assuming mock data from the Square Kilometre Array (SKA) observatory. We start by describing our mock data before showing the results of the MCMC chains with a specific focus on cosmology.

\subsection{Mock data for SKA}
We construct the mock data for the Square Kilometre Array following the method presented in Refs.~\cite{giri2019neutral,Giri:2022nxq}. This means we consider instrumental noise and cosmic variance contributions in a self-consistent way, while the effects from the foreground contamination are included via scale cuts applied to the mock data. This approach is similar to previous forecast studies for HERA and SKA \cite{Greig:2015qca,Greig:2017jdj,Park:2018ljd,Qin:2020pdx}.

We assume an antenna configuration mimicking the expected setup of the SKA-Low telescope \footnote{The antennae configuration used in this work is taken from \url{https://www.skao.int/sites/default/files/documents/d18-SKA-TEL-SKO-0000422_02_SKA1_LowConfigurationCoordinates-1.pdf}.}, which is currently under construction in Australia. While the SKA-Low is planned to consist of $\sim$ 512 antennae, the central core with a 1 km diameter will have densely filled antenna coverage. These dense, short baselines will sample the large-scale signal ($k\lesssim 1.5$ h/Mpc) quite well, providing the possibility to produce low-resolution images of the 21 cm signal during reionization and cosmic dawn \cite{giri2021measuring,bianco2021deep}.

Here we follow the procedure described in Ref.~\cite{giri2018optimal} to simulate the radio observations. The procedure relies on the open-source package {\sc Tools21cm} \cite{giri2020tools21cm}. We have chosen the exact observational setup as Ref.~\cite{Giri:2022nxq}, to which we refer for the exact values of the telescope parameters used in producing the mock observations. In order to obtain the error on the power spectrum due to instrumental effects, we use the Monte Carlo procedure given in Ref.~\cite{giri2019neutral}. We calculate this error assuming a total observation time of 1000 hours.

In Fig.~\ref{fig:SKAmock}, we show our mock power spectra separated into 12 different redshift bins. Only data points between $k=0.1-1.5$ h/Mpc are shown. All $k$-modes below and above this range are ignored, mimicking expected cuts due to foreground contamination and significant instrumental noise, respectively. The different redshift bins correspond to equal-spaced frequency steps of 10 MHz, guaranteeing that the data is not affected by light-cone effects \cite{Datta:2014fna,Ghara:2015yfa,Giri:2017nty}. The error bars shown for the mock observations appear small for low redshifts due to the range of the y-axis used in the figure to make the evolution of modelled lines clearly visible. We have compared these error values with the previous independent works \cite{Greig:2015qca,Qin:2020pdx}.

The coloured lines in Fig.~\ref{fig:SKAmock} show various predictions based on our default astrophysical model and varying the cosmological parameters within the {\tt Planck 2018} \cite{Planck:2018vyg} error bars (assuming a multivariate Gaussian distribution for all parameters except the sum of the neutrino masses which is fixed to 0.06 eV). The spread of the lines provides a first indication regarding the ability of the SKA-Low data to constrain cosmology. It needs to be emphasised, however, that there are many potential degeneracies between cosmological and the various astrophysical parameters which have to be accounted for before making more quantitative predictions. This is the goal of the following section.

\begin{table}
\renewcommand{\arraystretch}{1.2}
\centering
\begin{tabular}{l c c c}
\hline
\hline
Parameter name & Acronym & mock value  & prior range \\
\hline
Matter abundance & $\Omega_{m}$ & 0.315 & [0.27, 0.37]\\ 
Baryon abundance & $\Omega_{\rm b}$ & 0.049	& [0.04, 0.06]\\
Scalar amplitude & $A_{s}$ [$10^{-9}$]	& 2.07 & [1.0, 3.5]\\ 
Hubble parameter & $h$ & 0.68 &	[0.6, 0.8]\\ 
Spectral index &$n_{s}$ & 0.963 & [0.90, 1.02]\\
Neutrino masses & $\Sigma m_{\nu}$ [eV] & 0.06 & [0.0, 0.5]\\
\hline
Amplitude of $f_{*}$ & $f_{*,0}$ & 0.1 &	[0.01, 1.0]\\ 
Slope of $f_{*}$ & $\gamma_2$	& 0.5 &	[-0.79,-0.19]\\ 
Truncation of $f_{*}$ & $M_{t}$ [M$_{\odot}$/h] & $10^{8}$ & [3.2e6, 3.2e9]\\
Amplitude of $f_{\rm esc}$ & $f_{\rm esc,0}$ & 0.1 &	[0.01, 1.0]\\ 
Slope of $f_{\rm esc}$ & $\alpha_{\rm esc}$	& 0.5 &	[0,1]\\
Ampl. of X-ray rad.  &$f_X$ & 1.0 & [0.01,10]\\ 
Min X-ray energy & $E_{\rm min}$ [keV] & 0.5 & [0.3, 2.0]\\ 
\hline
\end{tabular}
\caption{Mock values and priors of all parameters varied during the MCMC inference.}
\label{tab:mcmc_params}
\end{table}

\subsection{MCMC results}
In order to forecast how well cosmological and astrophysical  parameters can be recovered with data from the SKA-low interferometer, we perform a MCMC inference analysis using the mock data presented above. We assume 13 free model parameters with default (mock) values and prior ranges provided in Table \ref{tab:mcmc_params}.

For the five standard cosmological parameters ($\Omega_m$, $\Omega_b$, $A_s$, $h_0$, and $n_s$) we assume default values from {\tt Planck} \citep{Planck:2018vyg}. The sum of the neutrino masses is set to the minimum value $\sum m_{\nu}=0.06$ assuming the normal mass hierarchy. Note that for simplicity we assume two of the three neutrino species to be massless. This approximation is acceptable since cosmological probes are sensitive to the sum of the neutrino masses.

The mock values of the astrophysical parameters are taken from \cite{Mirocha:2017aaa} and \citep{Park:2018ljd}. In particular, the stellar-to-halo mass ratio (determined by the parameters $f_{*,0}$, $\alpha_*$, and $M_t$) is calibrated to the luminosity functions from the Hubble Space Telescope \cite{Mirocha:2017aaa, Bouwens:2014fua}, the escape fraction ($f_{\rm esc,0}$, $\alpha_{\rm esc}$) is inspired by radiative-hydrodynamical simulations \cite{Kimm:2016kkj,Katz:2018xle}, and the X-ray flux parameters ($f_X$, $E_{\rm min}$) are obtained from Ref.~\cite{Das:2017fys}.

We run three MCMC chains with different assumptions regarding the theoretical uncertainty ($\sigma_{\rm th}$). The uncertainty is added in quadrature to the observational errors ($\sigma_{\rm obs}$), i.e., 
\begin{equation}
\sigma_{\rm tot}^2 = \sigma_{\rm obs}^2 + \sigma_{\rm th}^2,\hspace{0.5cm}\sigma_{\rm th}=c_{\rm th}P_{\rm obs},
\end{equation}
where $c_{\rm th}$ stands for the fractional theoretical error on the power spectrum.

For the first chain, we assume a $\sim 30$ percent theory error ($c_{\rm th}=0.31$) providing an optimistic estimate for the accuracy of current theoretical models. Note, however, that this level of precision has yet to be confirmed by cross-validation studies. For the second and third chain we reduce the theory error to 10 percent and 3 percent, respectively. We believe that none of the existing predictors for the 21 cm power spectrum currently are at this level of precision. However, a 10 percent accuracy level seems achievable by cross-validating and improving on currently available prediction methods. A 3 percent accuracy-level, on the other hand, may only be reached by developing next-generation methods to calculate the 21cm signal.

\begin{figure*} 
\centering
\includegraphics[width=0.95\textwidth,trim=0.2cm 0.1cm 0.5cm 0.5cm, clip]{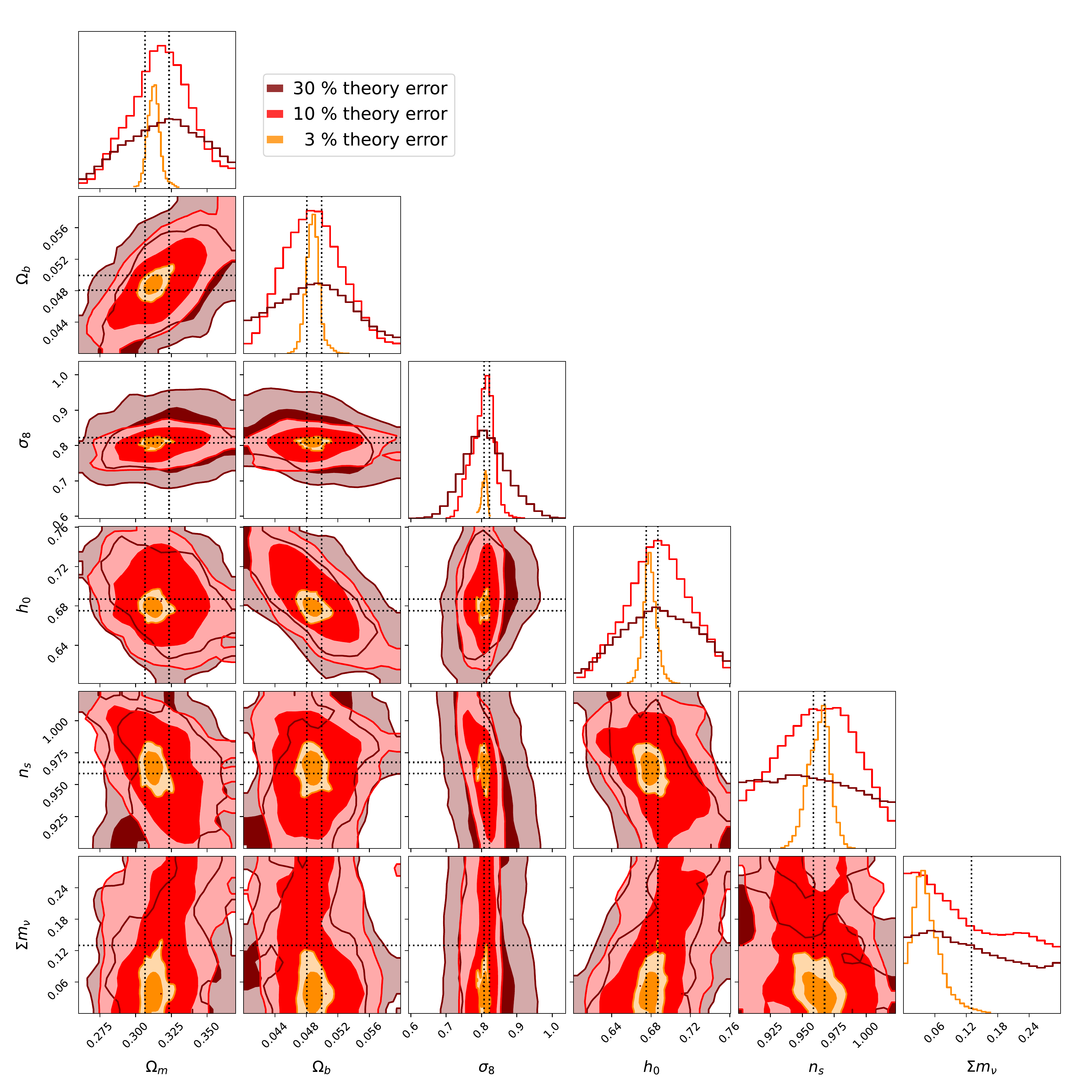}
\caption{Posterior contours of the cosmological parameters marginalising over all astrophysical nuisance parameters. Different colours correspond to different assumptions regarding the theory error (see caption). All posteriors are shown at the 68 and the 95 percent confidence level. For comparison we show the 68 percent errors from the {\tt Planck 2018} analysis \cite{Planck:2018vyg} as dotted lines.}
\label{fig:COSMOposteriors}
\end{figure*}

In Fig.~\ref{fig:COSMOposteriors} we show the posterior contours of the six cosmological parameters marginalising over all astrophysical parameters. The maroon, red, and orange colours correspond to the chains with a theory error of 30, 10, and 3 percent. All contours are shown at the 68 and 95 percent confidence level. The dotted lines indicate the 68 percent errors from the {\tt Planck 2018} analysis \cite{Planck:2018vyg} for comparison \footnote{The limits correspond to the TT, TE, EE + lowE data combination from the {\tt Planck 2018} analysis. For the sum of the neutrino masses, we calculate the 68 percent limit based on the published 95 percent constraint assuming a Gaussian error. See Table 2 and Eq. 58b in Ref.~\cite{Planck:2018vyg} for more information.}.

\begin{figure*} 
\centering
\includegraphics[width=0.95\textwidth,trim=0.2cm 0.1cm 0.5cm 0.5cm, clip]{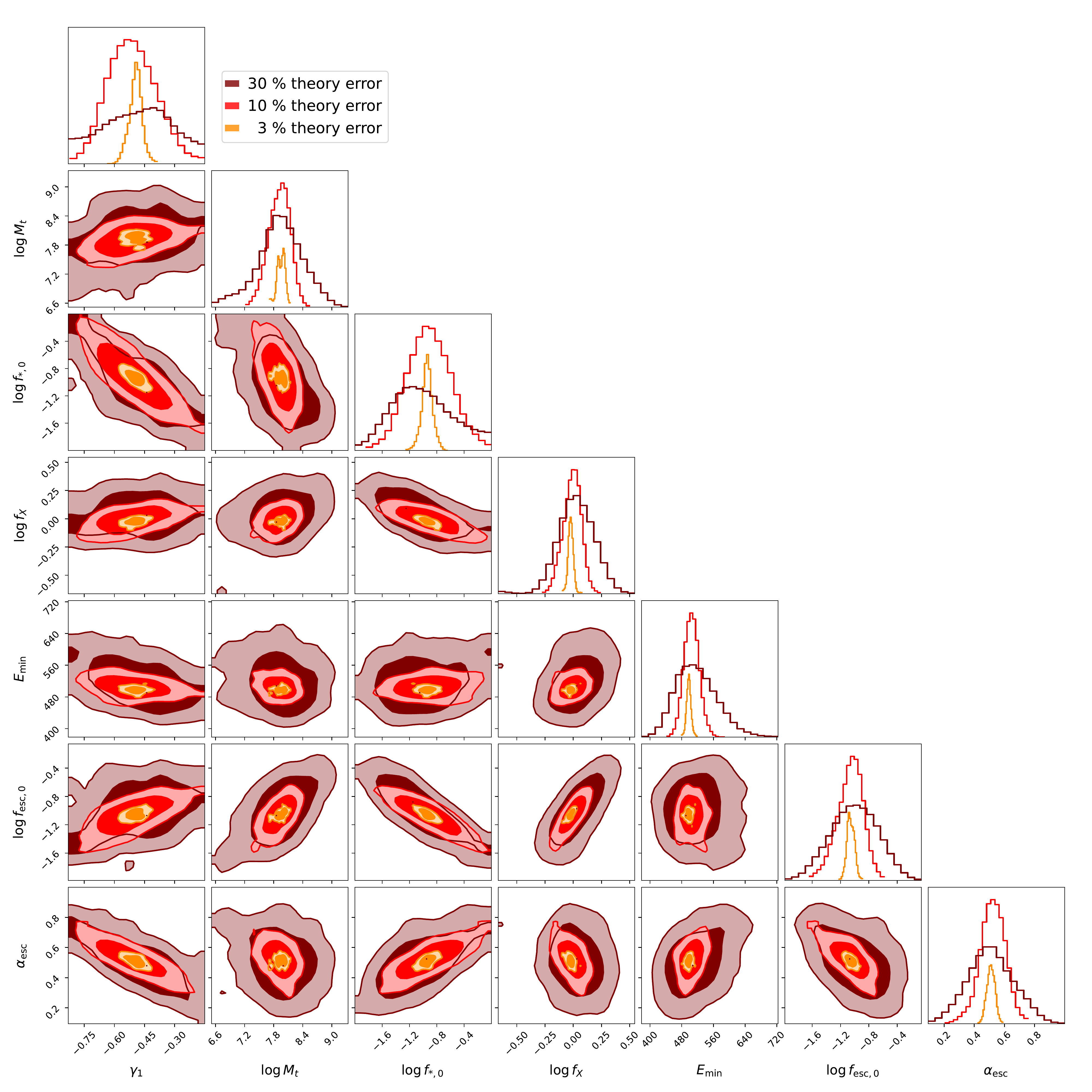}
\caption{Posterior contours of the astrophysical parameters marginalising over cosmology. Different colours correspond to different assumptions regarding the theory error (see caption). All contours are shown at the 68 and the 95 percent confidence level. }
\label{fig:ASTROposteriors}
\end{figure*}

The contours show that the 21-cm power spectrum can provide meaningful constraints on cosmology, even for the pessimistic case that the theory error cannot be pushed below 10 percent. Although the expected errors remain significantly larger than the ones from {\tt Planck}, it will be possible to validate our cosmological model with data from a very different redshift range, providing an important cross-test for the $\Lambda$CDM model.

For the case that the theory error can be reduced below the 10 percent level, the expected errors on the cosmological parameters shrink dramatically. As shown by the orange contours in Fig.~\ref{fig:COSMOposteriors}, which show the case of 3 percent theory error, the level of accuracy will be comparable to or even exceed the one from {\tt Planck}.

\begin{figure*} 
\centering
\includegraphics[width=0.48\textwidth,trim=0.2cm 0.1cm 0.5cm 0.5cm, clip]{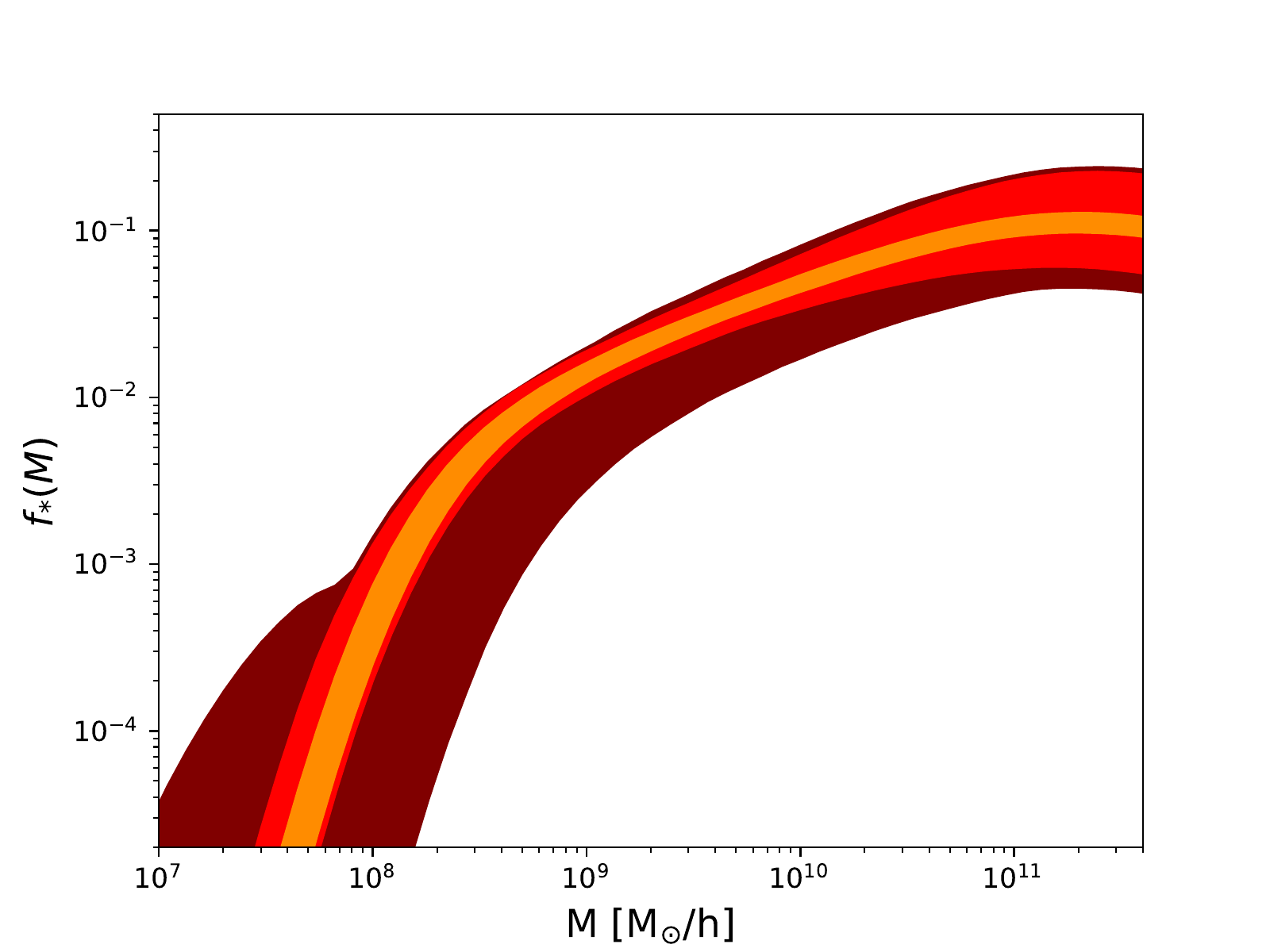}
\includegraphics[width=0.48\textwidth,trim=0.2cm 0.1cm 0.5cm 0.5cm, clip]{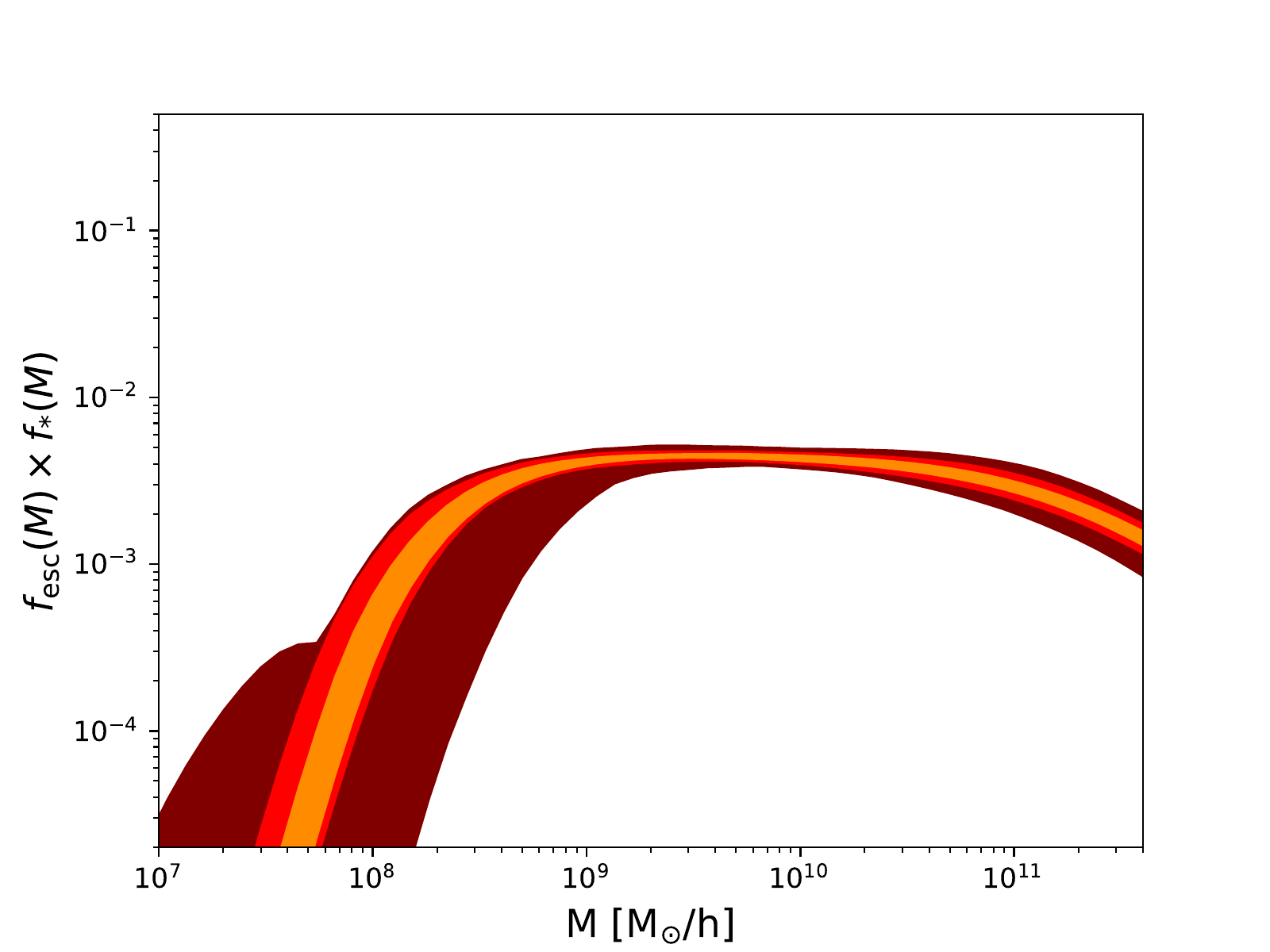}
\caption{Constraints on the stellar-to-halo mass ratio (left) and the product of the escape fraction with the stellar-to-halo mass ratio (right). The shaded bands correspond to the 68 percent confidence contours for the models with 30 percent (maroon), 10 percent (red), and 3 percent (orange) theory error.}
\label{fig:fstarfesc_post}
\end{figure*}

Most notably, the 21-cm power spectrum will be a powerful probe to constrain the neutrino sector, \cite[a fact pointed out at a more qualitative level by earlier work, see][]{McQuinn:2005hk,Pritchard:2008wy}. Our MCMC analysis with 3 percent theory error confirms that it will be possible to distinguish between the normal and inverted neutrino mass hierarchies (as the latter requires $\Sigma m_{\nu}>0.1$). Assuming the normal hierarchy in our mock setup, we are able to exclude the inverted hierarchy at more than 90 percent confidence. Furthermore, the unrealistic case of completely massless neutrinos is slightly disfavoured in our analysis, however, only at the 68 percent confidence level.

The other cosmological parameters can be constrained at a level equivalent to the {\tt Planck} data. This means that the 21-cm power spectrum from the epoch of reionization and cosmic dawn consists of a new and independent cosmological probe that can be used to stress-test the $\Lambda$CDM model. For example, it will be possible to provide an independent assessment of the Hubble ($H_0$) and $S_8$ tensions between the early-time and late-time cosmological probes. This is particularly interesting as the epoch of reionization and cosmic dawn lies in-between the CMB and low-redshift observables.

In Fig~\ref{fig:ASTROposteriors}, we show the posterior contours from the same chains as before, but this time plotting the seven astrophysical parameters marginalising over cosmology. Not surprisingly, the uncertainties on the astrophysical parameters decrease significantly when going from a 30 percent (maroon) to a 10 percent (red) and a 3 percent (orange) theory error.

A closer inspection of Fig.~\ref{fig:ASTROposteriors} reveals the many degeneracies between the different astrophysical parameters, most notably between $f_{*,0}$, $\gamma_1$, $f_{\rm esc,0}$, and $\alpha_{\rm esc}$. This is of course not surprising as the reionization signal is proportional to the product of the stellar-to-halo mass ratio and the escape fraction. However, it is worth noticing that these degeneracies are mainly visible in maroon and red but not so much in orange contours. This means that it will be possible to constrain $f_{*}(M)$ and $f_{\rm esc}(M)$ individually, provided the theory error can be pushed down to the 3 percent level. As a consequence, it will be possible to obtain stringent constraints on astrophysical quantities related to star formation, photon emission, and spectral source properties with the 21 cm power spectrum from SKA-Low.

Instead of showing the posterior contours at the level of individual parameters, we illustrate in Fig.~\ref{fig:fstarfesc_post} the constraints on the stellar-to-halo ratio and the escape fraction as a function of halo mass. The 68 percent confidence ranges of the two functions $f_*(M)$ and $f_{\rm esc}(M)\times f_*(M)$ from our MCMC analysis are plotted in the left- and right-hand panel, respectively. Different colours represent the assumed theory error of 30, 10, and 3 percent (maroon, red, orange) in agreement with the previous figures.

From Fig.~\ref{fig:fstarfesc_post} it becomes obvious that the 21-cm power spectrum of the SKA-Low telescope will allow putting stringent constraints on the galaxy-halo connection and the radiation properties of sources at redshifts $z\sim 6-20$. This information will be valuable for a better understanding of the first luminous sources and the advent of galaxy formation.

\section{Conclusions}\label{sec:conclusions}
Future radio interferometers like the Square Kilometre Array (SKA) telescope will observe the 21-cm signal at the epoch of reionization and cosmic dawn. From the theoretical side, the expected signal depends on the details of the source properties, such as their distribution and spectral flux. Furthermore, it is sensitive to cosmology as well as potential extensions of $\Lambda$CDM regarding the dark matter, dark energy, or early universe effects. From a statistical perspective, the problem contains a large number of free model parameters that have to be varied simultaneously during a Bayesian inference process. As a consequence, fast models are required in order to constrain the relevant model parameters with the upcoming data.

In this paper, we present a new, very fast method to calculate the 21-cm global signal and power spectrum at the epoch of reionization and cosmic dawn. The model ({\tt HMreio}) is based on the halo model of the cosmic dawn \cite{Schneider:2020xmf} combined with the excursion set bubble model from \citet{Furlanetto:2004nh} including an empirical factor to account for bubble overlap. We validate the {\tt HMreio} model by comparing it to the semi-numerical code {\tt 21cmFAST} \cite{Mesinger:2011aaa,Murray2020aaa}. Both models agree reasonably well for most redshifts and wave modes $k\lesssim 1$ h/Mpc as shown in Fig.~\ref{fig:comparison2}.

As a first application of the new model, we carry out a forecast analysis assuming mock data from a 1000 hours observing run with SKA-Low. The data includes the expected antenna configuration, collecting area, declination, system temperature, and bandwidth to obtain the survey specific instrumental noise and sample variance following the method of Ref.~\citep{Giri:2022nxq}. We also account for forground contamination via the scale-cut $k>0.1$ h/Mpc. The resulting mock data is shown in Fig.~\ref{fig:SKAmock}.

We carry out 3 different MCMC runs assuming an additional theoretical error of 30, 10, and 3 percent. A 30 percent theory error represents a pessimistic estimate of the precision level achievable with current codes. Obtaining a 10 and 3 percent theory error, on the other hand, will likely require substantial improvements or the development of new techniques.

The resulting posterior contours of the three cases are illustrated in Fig.~\ref{fig:COSMOposteriors} and \ref{fig:ASTROposteriors}. They show that the expected constraints on cosmological parameters will crucially depend on our ability to model the theoretical signal. While a 30 or 10 percent theory error will allow for first meaningful cross-checks of the $\Lambda$CDM model, a theory error below 3 percent will make it possible to obtain constraints exceeding the current {\tt Planck} limits from the CMB. Most notably, the sum of the neutrino masses will be very strongly constrained with data from {\tt SKA-Low}. Assuming a normal mass hierarchy as the truth, it will be possible to rule out the inverted hierarchy at 90 percent confidence. Furthermore, it will be possible to disfavour the non-physical case of massless neutrinos at  the 68 percent confidence level.

Though we focused on the 21-cm global signal and the power spectrum, the SKA-Low is expected to measure higher-order statistics, such as the 21-cm bispectrum \cite{majumdar2018quantifying,giri2019position,watkinson201921}. These higher-order statistics will contain more information and therefore have better constraining power \cite[e.g.][]{watkinson2022epoch}. Thus, the forecast study performed here gives a conservative estimate of the power of SKA-Low observations to constrain the cosmology.

Next to cosmology, the {\tt SKA-Low} telescope will allow us to learn more about various astrophysical relations. For example, it will be possible to strongly constrain the stellar-to-halo mass ratio down to halo masses below $10^8$ M$_{\odot}$/h. The same is true for the escape fraction of ionizing photons. Both relations will allow us to better understand the formation process of the first stars and galaxies in the universe.

We hope that the findings of this paper will provide further motivation to improve the modelling precision of the 21-cm signal at the epoch of reionization and cosmic dawn. This consists of a formidable challenge for the community of 21 cm astrophysics and cosmology.

\begin{acknowledgments}
This work is supported by the Swiss National Science Foundation (SNF) via the grant PCEFP2\_181157. Nordita is supported in part by NordForsk.
\end{acknowledgments}



\bibliography{HMreio}

\end{document}